\documentclass{article}
\usepackage{usenix,epsfig,endnotes}
\usepackage{amsmath}
\usepackage{amssymb}
\usepackage{graphicx}
\usepackage{epstopdf}
\usepackage{url}
\usepackage[letterpaper, margin=1in]{geometry}

\usepackage{subcaption}
\usepackage{booktabs}
\usepackage[table,xcdraw]{xcolor}
\usepackage{arydshln}
\usepackage{multirow}

\usepackage{pifont}
\newcommand{\cmark}{\ding{51}}%
\newcommand{\xmark}{\ding{55}}%
\usepackage[flushleft]{threeparttable}
\usepackage[underline=true]{pgf-umlsd}
\usetikzlibrary{calc}
\usepackage{listings, color}

\definecolor{gray}{rgb}{0.4,0.4,0.4}
\definecolor{darkblue}{rgb}{0.0,0.0,0.6}
\definecolor{cyan}{rgb}{0.0,0.6,0.6}

\lstdefinestyle{listXML}{language=XML, extendedchars=true,  belowcaptionskip=5pt, xleftmargin=1.8em, xrightmargin=0.5em, numbers=left, numberstyle=\small\ttfamily\bf, frame=single, breaklines=true, breakatwhitespace=true, breakindent=0pt, emph={}, emphstyle=\color{red}, basicstyle=\small\ttfamily, columns=fullflexible, showstringspaces=false, commentstyle=\color{gray}\upshape,
morestring=[b]",
morecomment=[s]{<?}{?>},
morecomment=[s][\color{orange}]{<!--}{-->},
keywordstyle=\color{cyan},
stringstyle=\color{black},
tagstyle=\color{darkblue},
morekeywords={xmlns,version,type}
}

\lstdefinestyle{customc}{
  belowcaptionskip=1\baselineskip,
  belowcaptionskip=5pt, xleftmargin=1.8em, xrightmargin=0.5em, numbers=left, numberstyle=\small\ttfamily\bf,
  breaklines=true, breakatwhitespace=true, breakindent=0pt, emph={}, emphstyle=\color{red}, basicstyle=\small\ttfamily, columns=fullflexible, showstringspaces=false, commentstyle=\color{gray}\upshape,
  breaklines=true,
  extendedchars=true,
  frame=single,
  morecomment=[s][\color{orange}]{/*}{*/},
  xleftmargin=\parindent,
  language=Java,
  showstringspaces=false,
  basicstyle=\footnotesize\ttfamily,
  keywordstyle=\bfseries\color{green!40!black},
  commentstyle=\itshape\color{purple!40!black},
  stringstyle=\color{orange},
  keywordstyle=\color{cyan},
stringstyle=\color{black},
tagstyle=\color{darkblue},
morekeywords={xmlns,version,type}
}
\usepackage{xtab, afterpage}
\usepackage{multicol}

\makeatletter
\def\@copyrightspace{\relax}
\makeatother

\begin{document}
%
% --- Author Metadata here ---
%\conferenceinfo{WOODSTOCK}{'97 El Paso, Texas USA}
%\CopyrightYear{2007} % Allows default copyright year (20XX) to be over-ridden - IF NEED BE.
%\crdata{0-12345-67-8/90/01}  % Allows default copyright data (0-89791-88-6/97/05) to be over-ridden - IF NEED BE.
% --- End of Author Metadata ---

\title{We Were Wrong: Ambient Sensors Do Not Provide Proximity Evidence for NFC Transactions}

\author{Raja Naeem Akram\thanks{r.n.akram@rhul.ac.uk}\qquad 
Iakovos Gurulian \thanks{Iakovos.Gurulian.2014@live.rhul.ac.uk} \qquad
Carlton Shepherd  \thanks{Carlton.Shepherd.2014@live.rhul.ac.uk} \\
Konstantinos Markantonakis \thanks{k.markantonakis@rhul.ac.uk} \qquad
Keith Mayes \thanks{keith.mayes@rhul.ac.uk}\\
Information Security Group, Royal Holloway, University of London.\\ 
Egham, United Kingdom. 
}

\maketitle
% Use the following at camera-ready time to suppress page numbers.
% Comment it out when you first submit the paper for review.
\thispagestyle{empty}

\begin{abstract}
Near Field Communication (NFC) has enabled mobile phones to emulate contactless smart cards. Similar to contactless smart cards, they are also susceptible to relay attacks.  To counter these, a number of methods have been proposed that rely primarily on ambient sensors as a proximity detection mechanism (also known as an anti-relay mechanism). In this paper, we, for the first time in academic literature, empirically evaluate a comprehensive set of ambient sensors for their effectiveness as a proximity detection mechanism.  We selected 15 out of a total of 17 sensors available via the Google Android platform for evaluation, with the other two sensors unavailable on widely-used handsets.  In existing academic literature, only 5 sensors have been proposed with positive results as a potential proximity detection mechanism.  Each sensor, where feasible, was used to record the measurements of 1000 contactless transactions at four different physical locations. A total of 252 random users, random sample of the university student population, were involved during the field trails. The analysis of these transactions provides an empirical foundation to categorically answer whether ambient sensors provide a strong proximity detection mechanism for security sensitive applications like banking, transport and high-security access control.  After careful analysis, we conclude that no single evaluated mobile ambient sensor is suitable for such critical applications in realistic deployment scenarios.  Lastly, we identify a number of potential avenues that may improve their effectiveness.
\end{abstract}

%\keywords{Contactless Transactions, Ambient Sensors, Proximity Detection, Anti-Relay Attack, NFC, Google Android, Mobile Phone, Smart Cards}

\section{Introduction}

Contactless smart cards are deployed in a multitude of applications, ranging from banking, transport and access control \cite{markantonakis2013secure}, and their awareness and acceptance is increasing, particularly in the banking sector \cite{UKCardsAnnualRep2015a}.  While the acceptance of the contactless cards is growing, mobile (contactless) payments are also gaining acceptance, particularly ``among early adopters and young age groups" \cite{UKCardsPayment2015a}.  In such mobile payments, Near Field Communication (NFC) provides the medium through which to emulate a contactless smart card; Deliotte projected in early 2015 that five percent of 600-650 million NFC-enabled mobile phones will be used at least once a month in 2015\footnote{Actual figures for 2015 were not available at the time of writing this paper. However, early forecasts for the global market of mobile payment for 2016 and beyond are available at http://www.nfcworld.com/technology/forecast/} to make a contactless payment \cite{Deloitte2015}.  We can reasonably assume, therefore, that mobile payments will be a significant payment medium in the future, potentially overtaking contactless smart cards. Although the above discussion is limited to the financial industry, similar trends are being observed in other domains where contactless smart cards are deployed \cite{VeriFone2010}.

Contactless smart cards, however, are susceptible to relay attacks \cite{Hancke2009615,Hancke06,kfir2005picking}, as are NFC-enabled mobile phones \cite{FrancisHMM11,FrancisHMM10,madlmayr2008nfc,6482441}.  A relay attack is a passive man-in-the-middle attack in which an attacker extends the distance between a genuine payment terminal (point-of-service) and genuine contactless smart card (or NFC-enabled mobile device). This attack can enable a malicious user to access services for which the genuine user is eligible, such as paying for goods or accessing a building with physical access controls. 

Quantifying the number of fraudulent activities where relay attacks were employed (on both smart card and NFC mobile phones) is a challenging task. Evidence exists, however, that academic work dealing with attacks on smart cards have been adopted by real-world criminals \cite{Ferradi2015}.  In the domain of contactless smart cards, a potentially effective countermeasure has been distance bounding protocols \cite{Hancke:2005:RDB:1128018.1128472,trujillo2010poulidor}.  For NFC-enabled phones, anti-relay mechanisms -- at least in academic literature -- have comprised largely of ambient sensing (Section \ref{sec:AmbientSensorDeploymetinMobilePayment}). In this paper, we investigate the ambient sensors available through the Google Android platform and construct a test-bed environment (Section \ref{sec:FrameworkforEvaluationofAmbientSensorsEffectiveness}) to evaluate their effectiveness as proximity detection mechanism for NFC-based contactless transactions (Section \ref{sec:AmbientSensorEvaluation}).  Additionally, we evaluate each individual sensor in relation to their technical feasibility, report any challenges that were encountered during implementation (Section \ref{sec:ExperimentationandChallenges}), and investigate issues affecting usability and reliability.  The aim of this work is to provide empirical evidence of each ambient sensor's suitability as a proximity detection mechanism and investigating whether it can counter potential relay attacks effectively (Section \ref{AntiRelayMechanismandAmbientSensorsTheVerdict}).

The suitability of a proximity detection mechanism for critical applications, such as banking, transport and (high-security) access control -- the main focus of this paper -- is based on its ability to uniquely pair measurements taken from a payment terminal and a payment instrument (or mobile handset in this case).  This is to provide confidence that the two devices were truly in close proximity ($\approx$ 3cm) to each other.  This measurement pair should be unique in a manner such that no other measurements can be paired successfully with the terminal or payment instrument's measurements.  That is to say, for an effective proximity detection mechanism that can thwart relay attacks, the device measurements generated during a transaction should be unique to that transaction.  Given a pair of measurements from two devices, \emph{PT} and \emph{PI}, representing the measurement from the payment terminal and the payment instrument respectively, the mechanism (and associated ambient sensor) is considered effective if it can uniquely pair \emph{PT} with \emph{PI}. No other measurement, \emph{PT'} or \emph{PI'}, can have the potential to be paired successfully with either \emph{PT} and/or \emph{PI}, thus providing a strong proof of proximity.  If there is a high likelihood that \emph{PT'} or \emph{PI'}, taken at different time or location, can be paired with \emph{PT} and/or \emph{PI}, then it is difficult for an entity (a third party or a payment terminal) to ascertain which device(s) were in actual proximity to each other.  This gap, where a transaction cannot be identified to have a unique measurement pair, is what a malicious user may exploit to mount a relay attack.   

In this paper, we continue to refer to contactless mobile payments due to the associated financial repercussions, and the attention this may attract from malicious actors.  However, the discussion in this paper is equally relevant towards the deployment of contactless mobile-based solutions in other domains, such as transport and access control.

There are three primary contributions of this paper:
\begin{enumerate}
\item A test-bed architecture and implementation that can be used to evaluate various sensors on Google Android devices.
\item A data analysis framework and methodology for evaluating ambient sensor measurements.
\item An empirical evaluation of the effectiveness of ambient sensors as a proximity detection mechanism.  This evaluation provides a foundation towards deciding which sensors to deploy in a target environment.
\end{enumerate}

The implementation of the test-bed, data analysis and collected data sets are being made available online at \url{http://anonymised}.

\section{Ambient Sensing in Mobile Payments}
\label{sec:AmbientSensorDeploymetinMobilePayment}
In this section, we briefly describe mobile phone-based contactless payments, relay attacks and a generic architecture for deploying ambient sensing as proximity detection mechanism to counter relay attacks. 

\subsection{Contactless Mobile Devices and Relay Attacks}
\label{sub:ContactlessMobileHandsetsAndRelayAttacks}
Distance bounding protocols in smart cards are closely integrated with the physical layer \cite{Hancke:2005:RDB:1128018.1128472}.  Usually, the overall architecture and process is the same with minor differences, whether the intended purpose of the smart cards is financial payments, transportation, access control or another domain.  The major differences reside in the higher-level protocols and algorithms that implement the semantics of each application.  In this section we examine contactless transactions in relation to NFC-based mobile payments; however, the overall architecture is common across all deployment domains.  %In this example, the implementation and workings of the NFC would be the same whereas at application layer individual applications can deploy their own algorithms/protocols. (CS: Repetition)

\begin{figure}[ht]
	\centering
		\includegraphics[width=0.95\columnwidth]{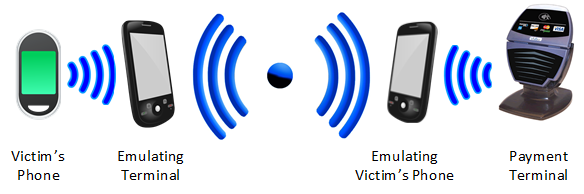}
	\caption{Overview of a Relay Attack}
	\label{fig:relay_attack}
\end{figure}

\begin{figure*}[ht]
	\centering
		\includegraphics[width=0.60\textwidth]{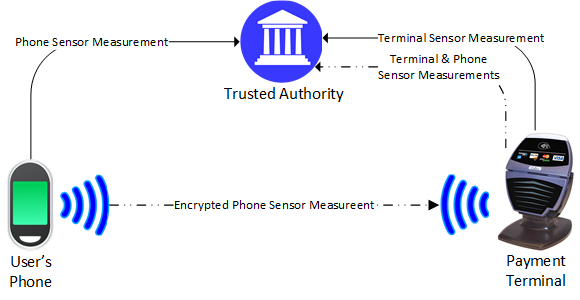}
	\caption{Generic Deployments of Ambient Sensors as Proximity Detection Mechanism}
	\label{fig:AmbientSensorDeploymentOverview}
\end{figure*}

When a user brings their NFC-enabled mobile phone close to a contactless payment terminal, the NFC in the mobile handset enters the radio frequency range around a payment terminal through which it can initiate a dialogue.  During a contactless transaction, physical contact is not necessary and, in many cases, a second factor of authentication, e.g.\ biometrics or Personal Identification Number (PIN), is not mandatory \cite{EVM2015-ContactlessArchitectureReq}.  As a result, it is difficult to ascertain that an authorised user brought the mobile device near the terminal via a relay.  It should be noted, however, that the use of a PIN or biometric may not counter a relay attack effectively in certain situations (notably the Mafia fraud attack \cite{Cremers2012}).

In a relay attack, shown in Figure \ref{fig:relay_attack}, an attacker must emulate a malicious payment terminal to a genuine user and a masquerading payment instrument (mobile phone) to a genuine payment terminal.  The goal of the malicious actor is to extend the physical distance of the communication channel between the victim's mobile phone and the payment terminal -- relaying each messages across this extended distance.  The attacker has the potential to pay for services using the victim's account if the attacker is able to relay these messages successfully without detection.

With contactless smart cards, relay attacks are countered using distance bounding protocols \cite{rasmussen2010realization} and variants of such \cite{Hancke2009615}; this is still an active research domain, with new attacks and countermeasures emerging \cite{Hancke:2008:ATD:1352533.1352566, Cremers2012, boureanu2014towards}.  The aim of distance bounding protocols is to enable a payment terminal (verifier) to enforce an upper bound on its physical distance to a smart card (prover).  The enforcement of physical distance is achieved through periodic challenge-response exchange and timing the process.  If the response is received within a predefined time (threshold), the verifier establishes that the prover is within a specified physical distance.  In a relay attack, provers may be either malicious or genuine; verifiers, however, are considered honest.  If verifiers are malicious and try to participate in a relay attack, then they are effectively attacking themselves, thus defeating the purpose of the attack. There is a possibility that a terminal might play a unwitting victim, but so long as we do not consider them to be physically modifying the (certified) payment terminals, then an anti-relay mechanism can succeed.

A substantial portion of the work surrounding relay-attack countermeasures for contactless smart cards relates to distance bounding protocols \cite{DrimerM07,Francillon11}.  However, these may not be feasible for NFC-enabled phones -- at the current state-of-the-art -- due to their requirement of high time-delay sensitivity and specialised hardware \cite{Coskun2013,Halevi2012}.

In the domain of NFC-based contactless mobile transactions, several methods have been proposed to provide some notion of proximity detection, most of which utilise environmental (ambient) sensors present on modern mobile handsets (for related work see Section \ref{sec:RelatedWork}).  In the following section, we discuss how ambient sensors have been proposed to counter relay attacks in NFC-based mobile contactless transactions. 

\subsection{Ambient Sensors for Proximity Detection}
\label{sec:AmbientSensorsforProximityDetection}
An ambient sensor measures a particular physical property of its immediate surroundings, such as temperature, light and sound.  Modern smartphones and tablets, are equipped with one or more of these sensors.  The physical environment surrounding a smartphone (or a payment terminal) can potentially provide a rich set of attributes that might be unique to that location -- the sound and lighting of a quiet, brightly-lit room, for example -- and such information might be useful to implement proximity detection between two interacting devices.  

In this section, we discuss a generic approach for deploying ambient sensing as a proximity detection mechanism that can be used in the context of mobile payments. Figure \ref{fig:AmbientSensorDeploymentOverview} illustrates the entities involved in this process. Variations of this approach are discussed below.

\begin{enumerate}
\item \textbf{Independent Reporting}.  In this scenario, depicted as solid lines in Figure \ref{fig:AmbientSensorDeploymentOverview}, both the smartphone and payment terminal collect sensor measurements independent of each other and transmit these to a trusted authority.  This authority compares the sensor measurements, based on some predefined comparison algorithm with set margins of error (threshold), and decides whether the two devices were are in proximity of each other.
\item \textbf{Payment Terminal Dependent Reporting}.  This setup, depicted as double-dot-dash line in Figure \ref{fig:AmbientSensorDeploymentOverview}, involves the smartphone encrypting the sensor measurements with a shared key (between smart phone and trusted authority) and transmitting the encrypted message to the payment terminal.  The payment terminal sends its own measurements and the smartphone's to the trusted authority for comparison.
\item \textbf{Payment Terminal (Localised) Evaluation}.  The smartphone transmits its own measurement to the payment terminal, which then compares it with its own measurements locally; the payment terminal then decides whether the smartphone is in proximity. 
\end{enumerate}

Regardless of how the user interacts with the payment terminal, e.g.\ touching or tapping it with their device, the overall deployment architecture falls under one of the above scenarios.  It can be observed that there is a potential for a fourth scenario to have a mobile phone perform the proximity analysis.  However, as we consider the payment terminal to be the (potentially) trusted device, as per the discussion in Section \ref{sub:ContactlessMobileHandsetsAndRelayAttacks} and the smartphone might be with a malicious entity, this scenario is not discussed.

\subsection{Related Work}
\label{sec:RelatedWork}
As noted previously, distance bounding protocols might not be suitable for mobile based contactless transactions.  In existing work, therefore, ambient sensors have been proposed as a strong proximity detection mechanism to counter relay attacks, which is discussed as below.

Drimer et al.\ \cite{DrimerM07} and Ma et al.\ \cite{6378376} showed how location-related data, namely using GPS (Global Positioning System), can be used to determine the proximity of two NFC mobile phones.  Ma et al.\ use a ten second window with location information collected every second, which was subsequently compared across various devices.  The authors report a high success rate in identifying the devices in close proximity to one another.

Halevi et al.\ \cite{Halevi2012} demonstrated the suitability of using ambient sound and light for proximity detection.  Here, the authors analysed the sensor measurements -- collected for 2 and 30 seconds duration for light and audio respectively -- using a range of similarity comparison algorithms.  Extensive experiments were performed in different physical locations, with a high success rate of detecting co-located devices.

Varshavsky et al.\ \cite{Varshavsky2007} based their proximity detection mechanism on the shared radio environment of devices -- the presence of WiFi access points and associated signal strengths -- using the application scenario of secure device pairing.  In this work, they considered this approach to produce low error rates, recommending it as a proximity detection mechanism.  While their paper did not focus on NFC-based mobile transactions, their techniques and methodology may still be applicable.

Urien et al.\ \cite{Urien201428} use ambient temperature with an elliptic curve-based RFID/NFC authentication protocol to determine whether two devices are co-located.  Using this method, they were successful in establishing a secure channel; the proposal combines the timing channels in RFID, traditionally used in distance bounding protocols, in conjunction with ambient temperature.  Their proposal, however, was not implemented and so there is no experimental data provided to evaluate its efficacy.

Mehrnezhad et al.\ \cite{mehrnezhad2014tap} proposed the use of an accelerometer to provide assurance that the mobile phone is within the vicinity of the payment terminal.  Their proposal requires the user to tap the payment terminal twice in succession, after which the sensor streams of the device and the payment terminal are compared for similarity.  It is difficult to deduce the total time it took to complete one transaction in its entirety, but the authors have provided a sensor recording time range of 0.6--1.5 seconds.

  \begin{table}[ht]
	\centering
	\caption{Related Work in Sensors as Anti-Relay Mechanism}
	\label{tab:RelatedWork}
    \resizebox{0.95\columnwidth}{!}{
\def\arraystretch{1.2}
			\begin{tabular}{@{}lccc@{}}
				\toprule
				\multirow{2}{*}{\textbf{Paper}} & \multirow{2}{*}{\textbf{Sensor}} 
                & \textbf{Sample} & \textbf{Contactless}\\
                && \textbf{Duration} & \textbf{Suitability}\\
                \midrule
    	 Ma et al.\ \cite{6378376} & GPS & 10 seconds & Possibly Not \\ [0.6ex]
         \hdashline 
        \multirow{2}{*}{Halevi et al.\ \cite{Halevi2012}} & Audio  & 30 seconds &  Possibly Not\\ 
        												 & Light  & 2 seconds &  Potentially \\ [0.8ex]
                                                         \hdashline
      	\multirow{2}{*}{Varshavsky et al.\ \cite{Varshavsky2007}} & WiFi & \multirow{2}{*}{1 second} & \multirow{2}{*}{Potentially} \\ 
        													     & (Radio Waves) &  &  \\ [0.6ex]
                                                                 \hdashline
    	Urien et al.\ \cite{Urien201428} & Temperature & N/A & - \\ [0.6ex]
        \hdashline
        Mehrnezhad et al.\ \cite{mehrnezhad2014tap} & Accelerometer & 0.6 to 1.5 Seconds & Potentially\\
    
		\bottomrule
			\end{tabular}
    }
\end{table}

We summarise the related work in Table \ref{tab:RelatedWork}, and use sensor sampling durations to determine whether a given approach is suitable for contactless NFC mobile phone transactions, namely in banking and transportation.  \emph{`Possibly not'} are those proposals whose sample duration is so large that they may not be adequate for mobile-based services that substitute contactless cards, whereas those whose sample range can be considered reasonable in certain applications are labelled as \emph{`Potentially'} in the Table \ref{tab:RelatedWork}. However, even schemes denoted as \emph{`Potentially'} may not be suitable for certain domains, such as banking or transport applications, where a strict upper-bound is often present in which to complete the entire transaction.  In these domains, the goal is to serve people as quickly as possible to maximise customer throughput, so time is critical in determining whether a transaction is successful and, indeed, permitted; an optimum transaction duration would be in milliseconds, rather than seconds.  We will return to the discussion about transaction duration limits later in Section \ref{sec:Test-Bed_Architecture}. 

Ambient sensing is also used in user-device authentication, key generation and establishment of secure channels.  These applications typically measure the environment for longer periods of time ($>$1 second) and, generally speaking, their primary goal is not proximity detection.  As such, we do not discuss them in this section.  

The field of ambient sensors for proximity detection in NFC-based mobile services is expanding, as illustrated by the number of proposals that currently exist.  In this paper we extend the discussion to a large set of ambient sensors.  We evaluate the suitability of sensors proposed currently and investigate a range of sensors not yet explored as an anti-relay mechanism in related work.  Table \ref{tab:availability} shows that we have undertaken a comprehensive evaluation of ambient sensors for proximity detection (fifteen in total).  In existing literature, only five of these sensors are proposed and evaluated as proximity detection mechanism (Table \ref{tab:RelatedWork}).  

\section{Framework for Evaluating Ambient Sensors}
\label{sec:FrameworkforEvaluationofAmbientSensorsEffectiveness}

In this section, we describe the test-bed that was developed to test, analyse and evaluate the effectiveness of using various ambient sensors as a proximity detection mechanism.  The results of the evaluation are presented in Section~\ref{sec:AmbientSensorEvaluation}.

\subsection{Evaluation Framework: Theory}
\label{sec:EvaluationFrameworkinTheory}
The aim of this work is to evaluate the effectiveness of ambient sensors as a proximity detection mechanism for high security mobile contactless transactions; to this end, we devised an evaluation framework discussed below.

% TODO (iakovos) maybe we should define what the term "transaction" means in this context.
% TODO (iakovos) The second sentence does not sound right.

For a particular ambient sensor, we collect sensor measurements on the Payment Terminal (PT) and the Payment Instrument (PI), and these measurements are referred to as a transaction pair comprising $PT$ and $PI$.  We collected a set (Equation \ref{dataset-fieldcollection}) of these transaction pairs, each containing $i$ transactions in total, by performing a field trial at four different locations on a university campus.

\begin{equation}
\{(PT_{1},~PI_{1}),(PT_{2},~PI_{2}), \dots, (PT_{i},~PI_{i})\}
\label{dataset-fieldcollection}
\end{equation}

For each transaction pair, we calculate the similarity between $PT$ and $PI$.  We refer to this as V, and thus our set (Equation \ref{dataset-fieldcollection}) now has three entries per transaction, as shown in Equation \ref{dataset-fieldcollection-variation}.

\begin{equation}
\{(PT_{1},~PI_{1},~V_1),(PT_{2},~PI_{2},~V_2), \dots, (PT_{i},~PI_{i},~V_i)\}
\label{dataset-fieldcollection-variation}
\end{equation}

From this set, we compute the Equal Error Rate (EER) threshold associated with each sensor by plotting the False Positive Rate (FPR) and False Negative Rate (FNR) curves, and locating the point at which they intersect.  %The notion of EERs is a standard metric in binary classification systems, such as in biometrics, where the rate of false positives is equal to false negatives.

In the context of our analysis, a False Positive is when a genuine pair\footnote{A genuine (value, transaction or pair) is a measurement(s) that we have collected during our trails and we have the evidence that they were taken from devices that were in close proximity to each other.} of $PT$ and $PI$ is not considered to be measured by devices that are in proximity to each other under a selected threshold ($V_x$).  Conversely, a False Negative is when a $PT_{i}$ matches to be in proximity to another $PI_{j}$ (where $i \neq j$) under the genuine and corresponding similarity value $V_i$ (threshold as threhold). 

%It is not equal in the above paragraph. If i and j are equal then they are basically representing genuine pair collected during the trail. 

For the FPR curve, we select $PT_j$ and $V_{j}$ and iterate through $n$ elements of the set, such that $n=1, 2, 3, \dots,~i$ except where $n = j$.  For each of the $n$ elements, if $V_j$ $<$ $V_n$ then the corresponding transaction (call it $T_n$) would result in False Positive. Rationale for this is, as we know that the T$_n$ is a genuine transaction, but because to the $V_j$ is smaller than the related $V_n$, the transaction will be declined if $V_j$ is the selected threshold.

For the FNR curve we select a value of $V_j$, then iterate through all the $n$ elements of the set such that $n=1, 2, 3, \dots,  i$ except $n = j$. For each of the $n^{th}$ element, we only take $PI_n$ and compute the similarity, $V_{jn}$, between $PT_j$ and $PI_n$. If the $V_{j}$ $>$ $V_{jn}$ then the result is a False Negative. The rationale behind this is that for a set of genuine transaction elements, ($PT_{j}$ and $V_{j}$), there exists another $PI$ that can be accepted -- other than the genuine value $PI_{j}$.   In this analysis, $V_{jn}$ is considered as the selected threshold.

In this section, some of the explanations were simplified for brevity and to avoid the nuances of sensor measurements and comparison techniques. In subsequent sections, we expand the discussion on the above points and provide a more detailed account of how we implemented the test-bed and evaluated the sensors data for each transaction. 

\subsection{Test-bed Architecture}
\label{sec:Test-Bed_Architecture}
\begin{figure}[ht]
	\centering
	\includegraphics[width=0.85\columnwidth]{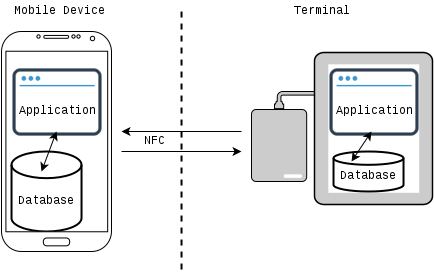}
	\caption{Test-bed Architecture}
	\label{fig:Architecture}
\end{figure}

Two applications were implemented and installed on pairs of Android devices, one as a payment terminal and the other as the payment instrument (mobile phone).
When the devices touch, an NFC connection is established and both begin recording data using a pre-specified sensor.  Upon successful completion of the measurements, each device stores the recorded data in a local database, shown in Figure~\ref{fig:Architecture}.

\begin{figure}[ht]
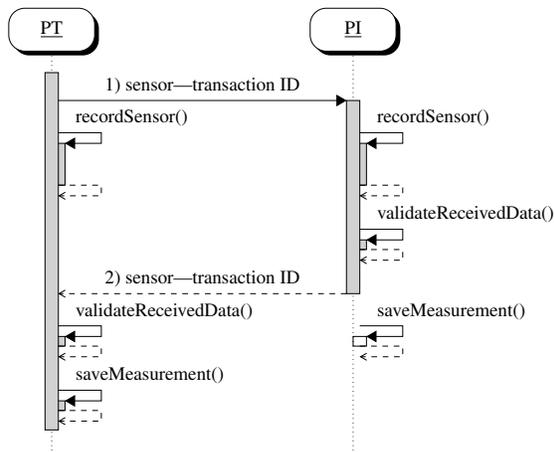

	\centering
	\resizebox{0.99\columnwidth}{!}{
		\begin{sequencediagram}

			\tikzstyle{inststyle}+=[rounded corners=3mm]
			\newthread{reader}{PT}
			\newinst[4]{card}{PI}
			\begin{call}{reader}{1) sensor|transaction ID}{card}{2) sensor|transaction ID}
				\begin{callself}{reader}{recordSensor()}{}
					\postlevel
				\end{callself}
				\prelevel\prelevel\prelevel
				\begin{callself}{card}{recordSensor()}{}
					\postlevel
				\end{callself}
				\begin{callself}{card}{validateReceivedData()}{}
				\end{callself}
			\end{call}

			\begin{callself}{card}{saveMeasurement()}{}
			\end{callself}
			\prelevel\prelevel
			\begin{callself}{reader}{validateReceivedData()}{}
			\end{callself}
			\begin{callself}{reader}{saveMeasurement()}{}
			\end{callself}

		\end{sequencediagram}
	}
	\caption{Measurement Recording Overview}
	\label{fig:diagram}
\end{figure}

A more detailed view of this process is depicted in Figure~\ref{fig:diagram}.  Bringing the two devices together causes the PT application to send a message to the PI, stating which sensor it uses in the transaction, along with a unique transaction ID\@.

After this message is received by the PI, both applications initiate the process to record a measurement, by storing values returned by a sensor.  The recording process lasts 500ms -- the maximum permitted time in which a contactless payment transaction should complete in accordance with the ``\textit{EMV Contactless Specifications for Payment Systems: Book A}" \cite{EVM2015-ContactlessArchitectureReq} and additional industrial specifications \cite{VISAMobileTicketing2013,MasterCard2014,VISA-TADG}. This time limit will be reduced gradually to 400ms from 2016 onward.  For transport-related transactions, the performance requirements are stricter, where transaction times should not exceed 300ms \cite{transport300,emms2014harvesting}.

After completing the measurements, the PI validates the data it received from the terminal (in message one), returning a completion or rejection message accordingly.  This is along with which sensor was used on the PI and the transaction ID received from the PT in message one.  This validation process ensures that the two devices were recording data from the same sensor.
%The terminal application can then ensure that it is in the same transaction as the mobile device, and if not, reject the measurement.
Finally, the PT verifies the data received from the PI, ensuring that the PI was measuring the same transaction, using the transaction ID and whether the two devices were recording from the same sensor. Upon validation, the devices save the measurements in their local databases.  In the event of any inconsistencies during the exchange, i.e.\ the devices recorded data for differing transaction IDs, then the measurement is rejected.  The database is designed to hold measurements for each transaction, which are used to analyse and evaluate the effectiveness of each sensor.

\subsection{Data Collection Framework}
\label{datacollectionsection}
To gauge the sensor variance in physical locations and how this influences proximity detection, we test each sensor in four different locations around the university: the lab, cafeteria, dining hall and library.  A field trail was conducted in each location with 252 participants supplied a varying number of transactions, with each providing a minimum of one transaction per sensor.

\begin{table}[h]
	\centering
	\caption{Sensor Availability}
	\label{tab:availability}
	\resizebox{0.95\columnwidth}{!}{
		\begin{threeparttable}
			\begin{tabular}{@{}lcccc@{}}
				\toprule
				\multicolumn{1}{c}{\textbf{Sensors}} & \textbf{Nexus 9 (1)} & \textbf{Nexus 9 (2)} & \textbf{Nexus 5} & \textbf{SGS5 mini} \\ \midrule
				\rowcolor[HTML]{EFEFEF}
				\multicolumn{5}{c}{\cellcolor[HTML]{EFEFEF}PT-PI Pair: Nexus 9 (1) $\rightarrow$ Nexus 9 (2)} \\
				\textbf{Accelerometer} & \cmark & \cmark & \cmark & \cmark \\
				\textbf{Bluetooth} & $\ast$ & $\ast$ & $\ast$ & $\ast$ \\
				\textbf{GRV$^\dag$} & \cmark & \cmark & $\ast$ & \cmark \\
				\textbf{GPS} & $\ast$ & $\ast$ & $\ast$ & $\ast$ \\
				\textbf{Gyroscope} & \cmark & \cmark & \cmark & \cmark \\
				\textbf{Magnetic Field} & \cmark & \cmark & \cmark & \cmark \\
				\textbf{Network Location} & \cmark & \cmark & \cmark & \cmark \\
				\textbf{Pressure} & \cmark & \cmark & \cmark & \xmark \\
				\textbf{Rotation Vector} & $\ast$ & $\ast$ & $\ast$ & $\ast$ \\
				\textbf{Sound} & \cmark & \cmark & \cmark & $\ast$ \\
				\textbf{WiFi} & $\ast$ & $\ast$ & $\ast$ & $\ast$ \\
				\multicolumn{5}{c}{\cellcolor[HTML]{EFEFEF}PT-PI Pair: SGS5 mini $\rightarrow$ Nexus 5} \\
				\textbf{Gravity} & $\circ$ & $\circ$ & \cmark & \cmark \\
				\textbf{Light} & $\ast$ & $\ast$ & \cmark & \cmark \\
				\textbf{Linear Acceleration} & $\circ$ & $\circ$ & \cmark & \cmark \\
				\textbf{Proximity} & \xmark & \xmark & \cmark & \cmark \\
				\multicolumn{5}{c}{\cellcolor[HTML]{EFEFEF}Unsupported} \\
				\textbf{Humidity} & \xmark & \xmark & \xmark & \xmark \\
				\textbf{Temperature} & \xmark & \xmark & \xmark & \xmark \\ \bottomrule
			\end{tabular}

			\begin{tablenotes}
			\item\cmark: Working properly.
				\xmark: Not present on device.
				$\ast$: Technical limitations.\\
				$\circ$: Only returning \textit{0s}.
			\item[\dag] Geomagnetic Rotation Vector
			\end{tablenotes}
		\end{threeparttable}

	}
\end{table}

Four devices were used in the experiments, forming two \textit{PT-PI device} pairs.
The first consisted of two Nexus 9 tablets, while the second pair comprised two Android smartphones: a Nexus 5, assuming the role of the payment terminal and a Samsung Galaxy S5 mini (SGS5 mini), acting as the payment instrument.  The availability of the sensors on each device can be found in Table~\ref{tab:availability}, along with which sensors comprised each of the two pairs discussed before.

Sensors with technical limitations -- Bluetooth, GPS, Rotation Vector and WiFi -- although present on the devices, returned no or very few data points ($>$99\% sensor failure\footnote{Detailed in Section \ref{sec:IndividualSensorResults} and Table \ref{tab:UsabilityandReliability}}) within the 500ms timeframe.  The specific limitations for each of these sensors are described in more detail in Appendix~\ref{sec:appendixB}.
Two sensors, the humidity and the temperature sensors, are relatively uncommon among Android devices and none of our tested devices contained them, thus we excluded them from this study.

A minimum of 1000 transactions -- measurement pairs for which both PT and PI have corresponding sensor data -- were recorded for each sensor.  700 measurements for each sensor were recorded in the Lab, with 100 in each of the remaining locations.

The Android operating system returns data captured by a sensor in time intervals set by the application.  To prevent unnecessary power consumption, the sensor values are returned by the operating system only when the values have altered.  The sound sensor (microphone), however, captures data in a continuous, uninterrupted stream.  In this instance, the applications convert the recorded amplitudes into sound pressure levels (in decibels) before storing the values in their respective databases.
With Bluetooth, data is sent from the operating system to the application every time a new Bluetooth device is discovered; with WiFi, this is once the device has completed scanning the presence of nearby access points.

On each device (PT and PI), each transaction generated a single database record containing the sensor values measured over 500ms.
The recorded sensor measurements were subsequently stored in the database in XML form. Every time sensor values were returned, a new child element was created containing the sequence ID of the measurement, the timestamp (initialised to zero at the start of the transaction), along with the recorded sensor measurements.  A sample XML record can be found in Listing~\ref{lst:xml}.
The data measurements -- whether a sensor returned one or multiple values per sample -- were saved in a generic format \textit{(e.g.\ data0, data1, etc.)} for convenience, and later assigned to their actual units during the analysis phase, in accordance with the Android documentation~\cite{sensorUnits}.
Details for each sensor can be found in Appendix~\ref{sec:AmbientSensors}.

The sequence ID of each measurement, the date and time the transaction occurred, the location in which it was captured, and the transaction ID were also stored in the database.

\begin{lstlisting}[style=listXML, caption={Sample XML Record}, label={lst:xml}, float, floatplacement=H]
<?xml version="1.0" encoding="UTF-8"?>
<measurements>
	....
	<measurement>
		<id>6</id>
		<timestamp>60</timestamp>
		<data0>-0.32935655</data0>
		<data1>6.2396774</data1>
		<data2>-7.558329</data2>
	</measurement>
	....
</measurements>
\end{lstlisting}

The transaction ID is a random 7-byte string generated by the terminal, used to assist in linking the measurements of each device to represent a single transaction.
%The loss of individual measurements typically occurs because the applications are built to reject any recorded data in case the NFC connection is lost, with the exception of the sound sensor, due to technical limitations, explained in section~\ref{sec:ImplementationChallenges}. (Repeats below)
Occasionally, the connection is disrupted when the mobile device has already replied to the terminal (message 2 in figure~\ref{fig:diagram}) while the terminal has not yet finished processing it.  As a result, the measurement is stored only on the mobile device; this occurs typically when the two devices are moved apart before the end of the transaction.  To counter this, the transaction ID is used in conjunction with the sequence ID to detect and exclude these measurements during the analysis phase.

%\begin{figure}[ht]
%	\centering
%	\includegraphics[width=0.75\columnwidth]{images/experiment.png}
%	\caption{Experiment}
%	\label{fig:Experiment}
%\end{figure}

After completing the experiments, the databases were extracted from the devices via USB and transferred to the PC running the analysis software.

\subsection{Data Analysis and Evaluation Criteria}
\label{sec:DataAnalysisandEvaluationCriteria}

After retrieving the databases from the terminal and mobile, the set of all transactions, $T$, was produced using the shared IDs generated during data collection.  Each transaction may be thought of as the set of PT and PI values, $PT_{i}$ and $PI_{i}$, with the same shared ID, i.e.\ $T_{i} = (PT_{i}, PI_{i})$. Whereas, each of the devices (PT and PI) will measure an array of data for each sensor at different time intervals. Therefore, $PT_{i}= {PT_{i1}, PT_{i2}, \dots, PT_{in}}$. Similarly, $PI_{i}$ is also a set of data vales collected by mobile during the $T_{i}$. A more detailed discussion on this is in Section \ref{sec:Data_Analaysis_Challenge} and depicted in Figure \ref{fig:workflow}.

A Python application was developed for analysing the transaction measurements from the application databases, employing the SciPy library \cite{scipy} for numerical computation.

%Sound equation
% Equations.  Please feel free to clean these up...
\begin{equation}
2r\footnote{$r$ represents the radius of Earth: 6371km} \arcsin  \left(\sqrt{\sin^2\frac{\phi_2 - \phi_1}{2}
+ \cos{\phi_1} \cos{\phi_2}\sin^2\frac{\lambda_2 - \lambda_1}{2}}\right)
\label{haversineeq}
\end{equation}

\begin{equation}
MAE(PT_{i}, PI_{i}) = \frac{1}{N}\sum_{j=0}^{N} | PT_{i,j} - PI_{i,j} |
\label{aadeq}
\end{equation}

\begin{equation}
corr(PT_{i}, PI_{i}) = \frac{covariance(PT_{i}, PI_{i})}{\sigma_{PT_{i}} \cdot \sigma_{PI_{i}}}
\label{correq}
\end{equation}

\begin{equation}
M = \sqrt{x^{2} + y^{2} + z^{2}}
\label{mageq}
\end{equation}

To compare $(PT_{i}, PI_{i})$, we measure the similarity distance between the two, which was measured differently according to sensor type. The reason behind this is how individual sensors are deployed in Android environment and number of variables returned in each data element during a transaction. To explain this further, during each transaction $T_i$, a PT and PI collect data from the selected sensor over a period of 500ms. During this interval, depending upon a sensor, a varying number of data elements is collected.  Each of these elements might contain one or more variables, for example, network location has two variables, accelerometer has three and temperature has only one variable per data element. Due to this, we devised three different methods of dealing with this diversity in the sensor data reporting.    
 
For network location, this was calculated using the Haversine formula (Eq.\ \ref{haversineeq}), which measures the geographic distance between two latitude and longitude pairs, $\{(\phi_1, \lambda_1), (\phi_2, \lambda_2)\}$.

For the remaining sensors, similarity was measured using the \emph{mean absolute error} (MAE, Eq.\ \ref{aadeq}) and \emph{correlation coefficient} (Eq.\ \ref{correq}), as used in \cite{mehrnezhad2014tap}, between the signals of $PT_{i}$ and $PI_{i}$, with each containing $N$ measurements. Over the series of $N$ data values per $PT_{i}$ and $PI_{i}$, we calculate the covariance to understand the overall change between the collected data (per transaction). Whereas, $\sigma_{PT_{i}}$ and $\sigma_{PI_{i}}$ is the mean of all the data points in $PT_i$ and $PI_i$, respectively.

Certain sensors -- the accelerometer, gyroscope, magnetic field, rotation vector and GRV sensors -- produce a vector of values comprising $x$, $y$ and $z$ components. In these instances, the vector magnitude (Eq.\ \ref{mageq}) was used as a general-purpose method for producing a single, combined value prior to computing the MAE and correlation coefficient.

\section{Challenges}
\label{sec:ExperimentationandChallenges}

In this section, we discuss the challenges encountered in implementing the test-bed, during data collection and in analysing and evaluating the collected data. The discussion is of value not only to the academic experimentation purposes but for actually testing a real deployment in the field -- based on ambient sensors to provide proximity detection.

\subsection{Implementation Challenges}
\label{sec:ImplementationChallenges}
A number of challenges were encountered during the implementation phase, particularly in relation to recording sufficient sensor values in a relatively short time period.

 The clocks of the two devices involved in a transaction cannot be expected to be completely synchronised, particularly in a real-world deployment.
		In order for the measurements on the devices to start as close to each other as possible, we perform the transaction validity checks after the recordings have finished. The validity checks involve verifying that both devices are measuring the same sensor and there is a unique transaction ID for the measurement. The recorded measurement is rejected at the end of the transaction in the event of the a validity check failure.

To capture the maximum possible amount of data, the applications were designed such that no threads, or other process-intense components, were used prior to the initialisation of the measurement recording (other than those started by the Android API automatically).
		The threads initiated by the Android API monitored any changes in the sensor values during a transaction.

\begin{figure*}
    \centering
    \begin{subfigure}[b]{0.31\textwidth}
        \includegraphics[width=\textwidth]{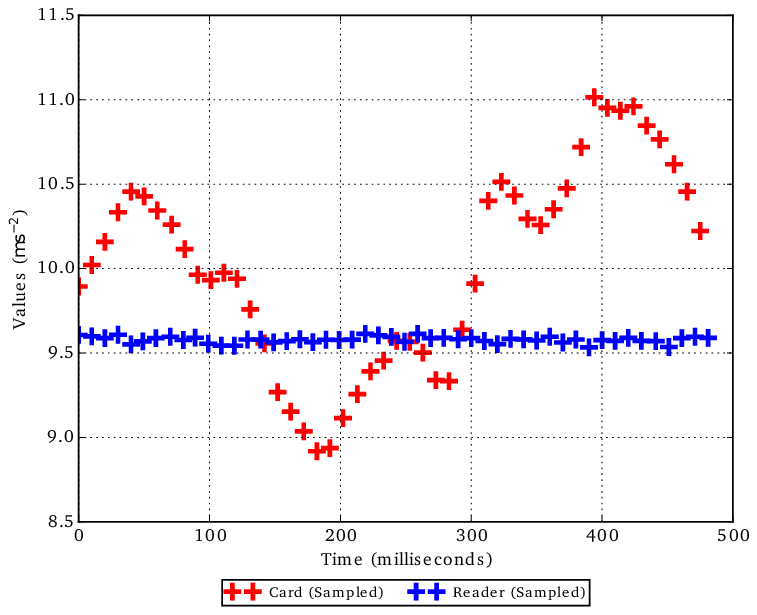}
        \caption{Step One}
        \label{fig:Step_One}
    \end{subfigure}
    ~ %add desired spacing between images, e. g. ~, \quad, \qquad, \hfill etc. 
    \begin{subfigure}[b]{0.31\textwidth}
        \includegraphics[width=\textwidth]{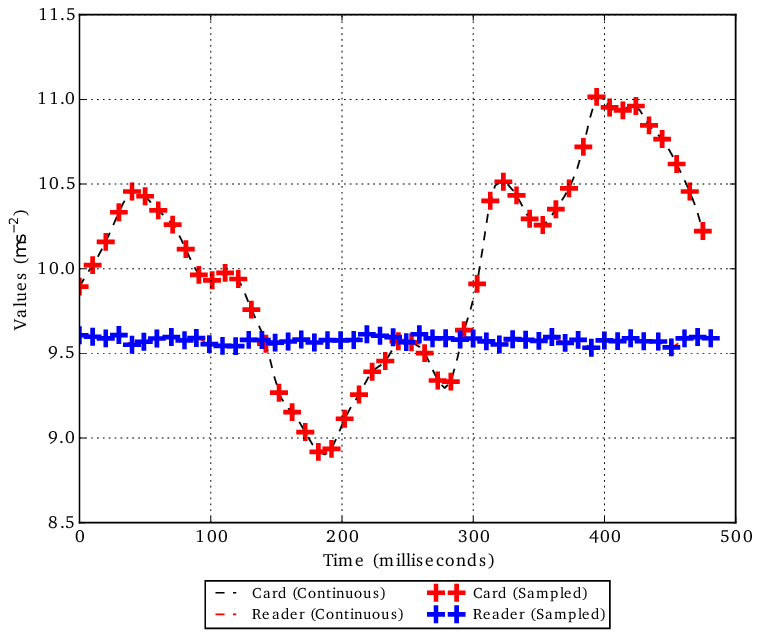}
        \caption{Step Two}
        \label{fig:Step_Two}
    \end{subfigure}
    ~ %add desired spacing between images, e. g. ~, \quad, \qquad, \hfill etc. 
    \begin{subfigure}[b]{0.31\textwidth}
        \includegraphics[width=\textwidth]{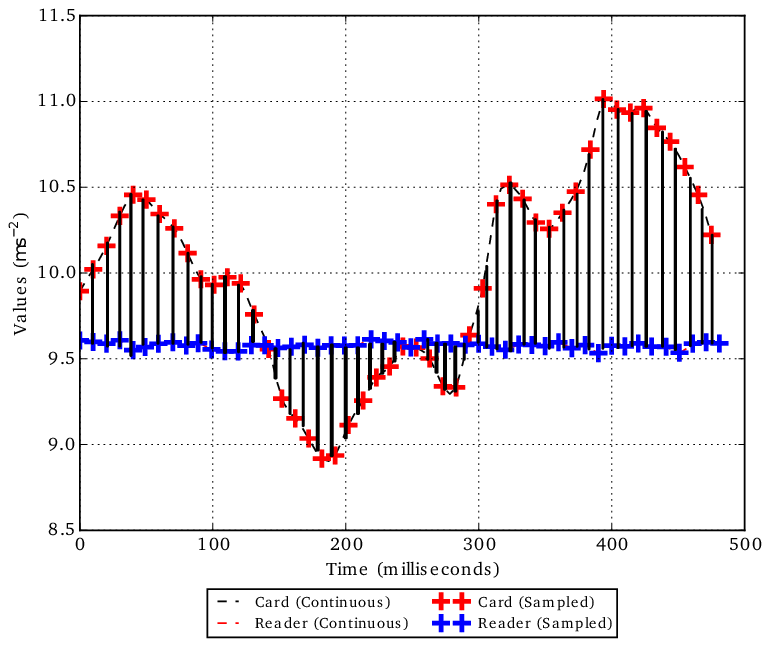}
        \caption{Step Three}
        \label{fig:Step_Three}
    \end{subfigure}
    \caption{Linear Interpolation to Mitigate the Effect of Missing and Inconsistent Sampling Rate -- An Example from Accelerometer-based Transactions}\label{fig:workflow}
\end{figure*}

The sound sensor's development was relatively challenging. Typically, to avoid any problems, the microphone should be accessed from within a separate thread, otherwise the NFC communication stops for as long as the microphone is being used. Initiating a thread prior to accessing the microphone had an impact in the performance of the application, leading to no data being recorded in 500ms. To overcome this issue, we measured 500ms chunks of sensor data in the devices, and not worrying about the loss of NFC connection during the transaction.

Ensuring that the recording of each sensor will not exceed 500ms was a high priority requirement that faced a number of challenges. A common way to run a task for a specific duration of time in Java is by using a timer inside a thread. Since the timer is initiated after the creation of a thread, which is a process intense procedure, there is no guarantee that the measurement will last for no more than 500ms since the initiation of the transaction. To overcome this issue, we track the system time on the terminal side when it sends the first message to the device and on the mobile device side, when it receives the first message from the terminal. A loop ensures that measurements after the 500ms timeframe will not be stored and the recording will terminate.

The \texttt{HostApduService} class was used for the implementation of the mobile device application, which is capable of emulating an NFC card on an Android handset, and is the basis of implementing a Host Card Emulation (HCE) service~\cite{hce}.
		HCE was introduced on Android 4.4 (API level 19), therefore the handset that acts as the PI in the experiment should be running Android 4.4 or later.

The data from the sensors was retrieved in time intervals of 10ms.
		Although the SGS5 mini device was capable of collecting data at a higher frequency (less than 1ms), the fastest common value across all our devices was 10ms.

\subsection{Data Collection Challenges}
Here, we discuss a selection of challenges faced during the data collection process.

Although the sampling rate for retrieving sensor values was set to 10ms, we encountered variance in the intervals that sensors returned data, as well as in the granularity of values provided across different devices.
		As mentioned previously, only when there is a significant change in the values recorded by a sensor are they returned to the application~\cite{AndroidAPIRef}.	As we are using a relatively small sampling rate (10ms), it is possible that there was no change in the values observed by the sensor, so no data was returned.
		Another reason relates to other processes, services and daemons running in the background that may also be causing slight delays to the frequency that an application is notified about sensor changes.
		During our experiments we ensured these delays were limited by uninstalling process intense applications and disabling any background services.  However, such precautions cannot be enforced in a real-world, wide-scale deployment by a general mobile application. However, a well embedded application into the Android platform with process priority and isolation might avoid it. 

Various sensors -- WiFi, GPS and Bluetooth -- were excluded from the study, since they were not capable of returning any useful data in the 500ms timeframe.
		We performed an initial limited experiment in which we only took 100 transactions per sensor prior to the data collection phase, and found that 100\% of these transactions from WiFi and GPS sensors contained little or no measurement data (i.e.\ at least one nearby WiFi access point or Bluetooth device). For Bluetooth, 99\% of the transactions contained little or no data points. As a result, they were excluded from the rest of our analysis.

\begin{table}[h]
	\centering
	\caption{Time Required for Receiving Results from Sensors}
	\label{tab:wifibt}
	\resizebox{0.95\columnwidth}{!}{
		\begin{tabular}{@{}lccc@{}}
			\toprule
			\multicolumn{1}{c}{\textbf{Sensor}} & \multicolumn{1}{c}{\textbf{Average time}} & \multicolumn{1}{c}{\textbf{Minimum time}} & \multicolumn{1}{c}{\textbf{Maximum time}} \\ \midrule
			\textbf{Bluetooth} & 962.7ms & 354ms & 1960ms \\
			\textbf{WiFi} & 3873.5ms & 3795ms & 3954ms \\ \bottomrule
		\end{tabular}
	}
\end{table}

        A separate application was built to measure the average time required for the WiFi and Bluetooth sensors to capture sufficient data.
		The GPS sensor's speed and accuracy are very dependent on the environment in which it is being tested.
		The results for 50 sensor measurements for the WiFi and the Bluetooth are presented in Table~\ref{tab:wifibt}.

\subsection{Data Analysis Challenges}
\label{sec:Data_Analaysis_Challenge}

During the data analysis, we faced a number of challenges.  Below is a list of some selected challenges that we had to overcome.

After acquiring the data from the PT and PI, we used linear interpolation to mitigate the effect of missing and inconsistent samples due to the unwanted variation in sampling rates (steps \textbf{1} and \textbf{2} of Figure \ref{fig:workflow}).

% \begin{figure*}
% \centering
% \includegraphics[width=\textwidth]{images/FlowsThreeGraph.pdf}
% \caption{Data analysis workflow for an accelerometer-based transaction}
% \label{fig:workflow}
% \end{figure*}

\begin{figure}[h]
\centering
\includegraphics[width=0.95\columnwidth]{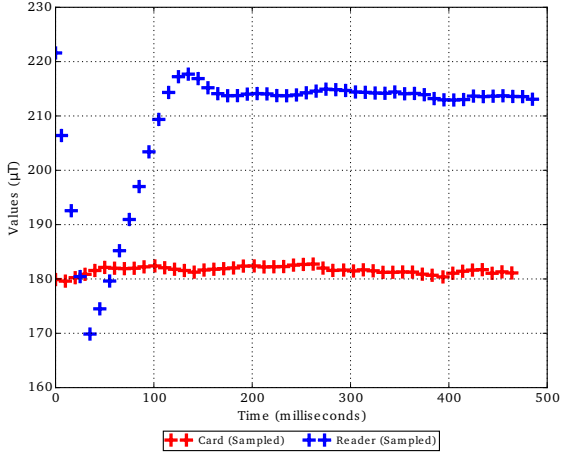}
\caption{Example Magnetic Field transaction illustrating differing finishing times for the PT and PI}
\label{fig:magfield}
\end{figure}

Data sampling rarely finished at 500ms precisely: Figure \ref{fig:magfield} illustrates that the reader's final sample occurs at $\approx$490ms, while the card finished at $\approx$460ms.  To counter this, interpolation was performed up to the maximum time period shared by both devices (460ms) and any samples beyond this point were discarded.  

 Correlation, however, is sensitive to this type of truncation: if relatively extreme values are discarded, such as the reader's measurements between 0-100ms in Figure \ref{fig:magfield}, then correlation may differ significantly. Thus, for computing correlation we avoid truncation.  

 Only one measurement was collected at maximum -- if any were collected at all -- when analysing network location for all transactions.  Since correlation is undefined for a single point, we calculated only the MAE for this particular modality.

For 43\% of gyroscope-based transactions (430 of 1000), a value of zero was measured for the entire transaction on the PT and/or PI, indicating that the device was not perturbed sufficiently to alter the sensor's resting value.  Correlation is undefined for a single point, so these transactions were discarded for the correlation analysis. This also suggests that gyroscope sensor might not be a suitable sensor for mobile based contactless transactions.  

 On one device (SGS5 mini), the light sensor returned values far more frequently than the other.  With the Nexus 5, only one light measurement was captured on average during the 500ms timeframe, while the SGS5 mini collected 48--50 measurements.  Rather than truncating a significant number of measurements, we assumed\footnote{Our assumption is based on the Android sensor API implementation and potentially the capability of the light sensor on Nexus 5. In which if the sensor value do not change, the Android API do not return any new values. For this reason, we assumed that for the complete duration of 500ms, the light sensor value didn't change from the first reported value.} that all of the Nexus 5 values lay on a straight line, with a gradient of zero, and calculated the MAE using this. Note that correlation is undefined in this case.

 The proximity sensor on both tested devices (Nexus 5 and SGS5 mini -- see Table \ref{tab:OptimumThresholds}) returned only a binary value, representing \emph{near} and \emph{far}, rather than the precise proximity of the device to an object in centimetres.  The Android documentation states that ``some'' devices provide a binary value rather than a real value \cite{androidposition}.  In our dataset, both devices reported \emph{far} for all transactions. The reasoning for this is because the proximity sensor is located on the front of the device and NFC being on the back of the devices. To tab two handset for NFC transaction, you have to touch the backsides of the devices with each other, which does not interact with proximity sensor as they are located at the front. Thus both devices reporting the values as ``far''.

\section{Ambient Sensor Evaluation}
\label{sec:AmbientSensorEvaluation}

%\subsection{Individual Location Analysis of Data}

In this section, we describe our methodology in detail to calculate False Positive Rate (FPR), False Negative Rate (FNR) and Equal Error Rate (EER). Furthermore, we present the results of our analysis for each individual sensor.  

\subsection{Calculating the FPR, FNR and EER}
As discussed in Section \ref{sec:EvaluationFrameworkinTheory} and
 \ref{sec:DataAnalysisandEvaluationCriteria}, each transaction, $T_i$, comprises a set of $PT_i$ and $PI_i$ measurements (Equation \ref{dataset-fieldcollection}).  From this data, we compute the $MAE(PT_i, PI_i)$ and $corr(PT_i, PI_i)$ for each successful transaction (collective we called them $V_i$ in Equation \ref{dataset-fieldcollection-variation}).   We calculate the FPR, FNR and EER of each sensor by testing the $MAE$ and $corr$ of the set of genuine transaction pairs\footnote{From trail data to be pairs from device that are in proximity to each other}, $(PT_i, PI_i)$, against the $MAE$ and $corr$ of unauthorised pairs $(PT_i, PI_j)$ using some threshold, $t$.  Ideally, $V_i(PT_i, PI_j) < t$ and $V_{ij}(PT_i, PI_j) > t$ for all possible pairs.  
 
The FPR and FNR are calculated using Equation \ref{eq:fnr}, where FP, FN, TP and TN represent the total number of False Positives, False Negatives, True Positives and True Negatives respectively for a given threshold.

\begin{equation}
FPR = \frac{FP}{FP + TN}
\qquad
FNR = \frac{FN}{FN + TP}
\label{eq:fnr}
\end{equation}

To elaborate upon this further, the pseudo-code for calculation of the FPR is given in Listing \ref{CalculatingFAR}:

\begin{lstlisting}[style=customc, caption={Calculating False Positive Rate},label=CalculatingFAR]
/* Pseudocode to calculate FPR data points */
int t = Tested_Threshold;
int maxint = Number_of_Transactions;
int FPS = 0; // False positives
int TNS = 0; // True negatives
for (int i = 1; i <= maxint; i++) {
  for (int j = 1; j <= maxint; j++) {
    if (i != j) {
      calculateMAE(PTi, Mj);
      if (MAE(PTi, PIj) < t)
         FPS++;
      else
         TNS++;
    }        
  }
}
double FPR = FPS / (FPS + TNS);
write(FPR);
\end{lstlisting}

The calculation of FPR based on $corr$ follows the same algorithm, only substituting $MAE$ with the $corr$ function.  To calculate the False Negative Rate (FNR), we implemented the algorithm described in Listing \ref{CalculatingFNR}.  As with FPR, substituting $MAE$ for $corr$ in Listing \ref{CalculatingFNR} allows us to generate an FNR curve based on correlation. 

\begin{lstlisting}[style=customc, caption={Calculating False Negative Rate},label=CalculatingFNR]
/* Pseudocode to calculate FNR data points */
int t = Test_Threshold;
int maxint = Number_of_Transactions;
int FNS = 0;  // False negatives
int TPS = 0;  // True positives
for (int i = 1; i <= maxint; i++) {
  calculateMAE(PTi, PIi)
  if (MAE(PTi, PIi) < t)
    TPS++;
  else
    FNS++;
}
double FNR = FNS / (FNS + TPS);
write(FNR);
\end{lstlisting}

\subsection{Individual Sensor Results}
\label{sec:IndividualSensorResults}

The aim of our evaluation is to investigate to what extent legitimate and illegitimate transactions can be identified using these similarity metrics.  For a transaction between two co-located devices, then $MAE(PT_{i}, PI_{i}) \approx 0$ and $corr(PT_{i}, PI_{i}) \approx 1$, while for a PT and PI device in differing locations, i.e.\ $(PT_{i}, PI_{j})$, the distance and correlation should be sufficiently large.  What is considered `sufficient' is determined through finding a suitable threshold, $t$, which permits all legitimate transactions while denying those which are illegitimate, i.e.\ $V_i(PT_{i}, PI_{i}) < t$ and $V_{ij}(PT_{i}, PI_{j}) > t$, as mentioned previously.  For each individual sensor, we aim to find an optimal value of $t$, along with its reliability.

\begin{figure}[ht]
	\centering
		\includegraphics[width=0.90\columnwidth]{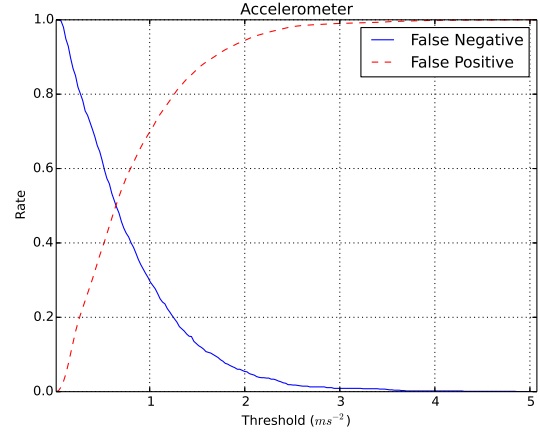}
	\caption{FPR and FNR Graph for an Accelerometer-based Transactions using $MAE(PT_i,PI_i)$}
	\label{fig:EERAmbientSensor}
\end{figure}

We generate FPR and FNR curves for $MAE$ and $corr$ for every sensor we were able to collect data.  The point of intersection for these curves provides an optimal threshold for $MAE$ and $corr$ based on its associated EER, i.e.\ the rate at which the acceptance and rejection errors are equal. Figure \ref{fig:EERAmbientSensor} shows the FPR and FNR curves for the accelerometer (based on $MAE$ calculations), along with the optimum threshold and EER\@. Appendix \ref{sec:AmbientSensors} includes the EER graphs for all of the tested sensors (except Accelerometer), based on both $MAE$ and $corr$. 

Figure \ref{fig:EERAmbientSensor} illustrates that the optimum threshold (based on the $MAE$) is 0.651, with an associated EER of 0.506.  A threshold of 0.651ms$^{-2}$, therefore, one can expect the proportion of false positives and false negatives to be 50.6\%.

\begin{figure}[ht]
	\centering
		\includegraphics[width=0.90\columnwidth]{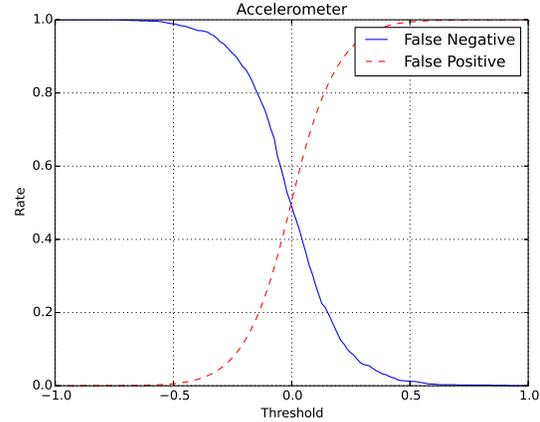}
	\caption{FPR and FNR Graph for Accelerometer-based Transactions using $corr(PT_i, PI_i)$}
	\label{fig:EERAmbientSensorcorr}
\end{figure}

Similarly, for correlation, Figure \ref{fig:EERAmbientSensorcorr} shows that the EER (0.523) produces an optimal threshold of 0.007.  If we use the $corr$ as the comparison criterion, the rate of false positives and negatives is approximately 52\%.  Table \ref{tab:OptimumThresholds} lists the optimum thresholds and associated EERs for each tested sensor.

\begin{table}[ht]
	\centering
	\caption{Optimum Thresholds and Associated Risk}
	\label{tab:OptimumThresholds}
	\resizebox{0.95\columnwidth}{!}{
    \begin{threeparttable}
			\begin{tabular}{@{}lcccc@{}}
				\toprule
				\multirow{2}{*}{\textbf{Sensors}} & \textbf{Optimum} & \multirow{2}{*}{\textbf{$EER_{MAE}$}} & \textbf{Optimum} & \multirow{2}{*}{\textbf{$EER_{corr}$}} \\ 
                								  & \textbf{Threshold$_{MAE}$} & & \textbf{Threshold$_{corr}$}& \\
                                                  \midrule
				\textbf{Accelerometer} & 0.651 & 0.506 & 0.007 & 0.523 \\
				\textbf{Bluetooth} & -- & -- & -- & -- \\
				\textbf{GRV} & 0.499 & 0.615 & 0.006 & 0.518 \\
				\textbf{GPS} & -- & -- & -- & -- \\
				\textbf{Gyroscope} & 0.520 & 0.504 & 0.007 & 0.752 \\
				\textbf{Magnetic Field} & 77.33 & 0.334 & 0.235 & 0.582 \\
				\textbf{Network Location} & 8.532 & 0.369 & N/A$^\ast$ & N/A \\
				\textbf{Pressure} & 2.787 & 0.270 & 0.329 & 0.645 \\
				\textbf{Rotation Vector} & -- & -- & -- & -- \\
				\textbf{Sound} & 10.19 & 0.395 & -0.087 & 0.524 \\
				\textbf{WiFi} & -- & -- & -- & -- \\
				\textbf{Gravity} & 9.5e-05 & 0.506 & 0.020 & 0.534 \\
				\textbf{Light} & 178.5 & 0.518 & 0.020 & 0.515 \\
				\textbf{Linear Acceleration} & 1.348 & 0.514 & 0.087 & 0.491 \\
				\textbf{Proximity} & N/A$^\dag$ & N/A & N/A & N/A \\
				\bottomrule
			\end{tabular}
            
            \begin{tablenotes}
            	\item $^\ast$Insufficient data to calculate correlation
                \item $^\dag$All transactions contained the same value for both devices.  Consequently, EER is undefined.
            \end{tablenotes}
            \end{threeparttable}
            }
\end{table}

In a wide-scale deployment of an ambient sensing proximity detection mechanism, a single threshold should be defined. The terminal (or third party) would store this threshold (Section \ref{sec:AmbientSensorsforProximityDetection}), and if the similarity of the terminal's and device's sensor readings was within this, then the transaction would be assumed to be legitimate, i.e.\ both devices in close proximity. However, setting a threshold of this nature invariably incurs some rate of false positives and false negatives.  The intersection of FPR and FNR provides us with the proportion of potentially malicious transactions might passing as genuine (false positives) and the proportion of genuine transactions being rejected (false negatives). The goal of a malicious entity would be to carry out relay attacks such that the sensor measurements at the terminal and mobile phone remained within the predefined threshold.  A threshold with a higher FPR provides a large working space to the attacker, whereas a higher FNR will reduce the usability of the scheme, potentially frustrating consumers by rejecting legitimate transactions.

Besides investigating the EERs of sensors and the effect this has on their suitability for NFC mobile services, we evaluate the reliability and potential usability of the selected sensors.   This analysis, shown in Table \ref{tab:UsabilityandReliability}, presents our findings regarding the proportion of failed transactions and sensor failures.

To collect 1000 transactions from 252 users (for each sensor), as explained in Section \ref{datacollectionsection}, we requested users to present the PI to the PT as many times as they preferred.  We established walk-in counters at four different locations of the university campus and students walking nearby by were invited to assist us in the trail.  We do not take any demographic data about the students as sensors are not used to identify a user, but to assure that two devices were in close proximity to each other during a transaction.  At times, transactions were not registered during this process, usually due to the user moving the handset away too quickly, and was the primary cause of transaction failures\footnote{These failed transactions were not included in the data analysis and results represented in Table 4, which is based on successful 1000 transactions.} (no shared measurements between the PT and PI) represented in Table \ref{tab:UsabilityandReliability}. The rate of sensor failures, in the same table, represents the situation when the transaction was successfully completed on both the PT and PI, but where one or both devices failed to record any data in the 500ms timeframe.  The percentage of transaction failures relates to the total transactions, while sensor failures is measured with respect to the number of successful transactions.  The transactions failure represents the difficulty in using the sensors by the user, while the sensor failure rates reflects their reliability.

\begin{table}[t]
	\centering
	\caption{Usability and Reliability Analysis}
	\label{tab:UsabilityandReliability}
	\resizebox{0.85\columnwidth}{!}{
			\begin{tabular}{@{}lccc@{}}
				\toprule
				\multirow{2}{*}{\textbf{Sensors}} & \textbf{Total} &\textbf{Transaction} & \textbf{Sensor} \\ 
                                                  & \textbf{Transactions}& \textbf{Failures} &  \textbf{Failures}\\
                								  \midrule
				\textbf{Accelerometer} & 1025 & 13 (1.26\%) & 0 (0\%) \\
				\textbf{Bluetooth} & 101 & 1 (0.99\%) & 99 (99\%) \\
				\textbf{GRV} & 1019 & 8 (0.78\%) & 0 (0\%) \\
				\textbf{GPS} & 101 & 1 (0.99\%) & 100 (100\%) \\
				\textbf{Gyroscope} & 1022 & 11 (1.07\%) & 0 (0\%) \\
				\textbf{Magnetic Field} & 1027 & 17 (1.65\%) & 0 (0\%) \\
				\textbf{Network Location} & 1053 & 15 (1.42\%) & 960 (96\%) \\
				\textbf{Pressure} & 1018 & 10 (0.98\%) & 0 (0\%)\\
				\textbf{Rotation Vector} & 1023 & 14 (1.36\%) & 0 (0\%)\\
				\textbf{Sound} & 1047 & 4 (0.38\%) & 0 (0\%) \\
				\textbf{WiFi} & 100 & 0 (0\%)&  100 (100\%)\\
				\textbf{Gravity} & 1165 & 143 (12.27\%) & 0  (0\%)\\
				\textbf{Light} & 1057 & 37 (3.50\%) & 0  (0\%)\\
				\textbf{Linear Acceleration} & 1175 & 159 (13.53\%) & 3 (0.3\%)\\
				\textbf{Proximity} & 1071 & 58 (5.41\%) & 0  (0\%)\\
				\bottomrule
			\end{tabular}
            }
\end{table}

\section{Outcome and Future Directions}
\label{AntiRelayMechanismandAmbientSensorsTheVerdict}

The results we present provide us with an empirical foundation for evaluating the suitability of various mobile sensors as a proximity detection mechanism for NFC-based mobile transactions.  As discussed in the previous section, the higher the EER, the greater the likelihood that an attack passes undetected and that a genuine transaction is rejected.  Based on our analysis, it is difficult to recommend any of the sensors (individually) for a high security deployment application, such as banking.  These sensors, however, might be appropriate for low-security access control, but we recommend that a thorough analysis of the sensors and their performance in the chosen domain is performed prior to deployment.

One potential reason that related research in this domain has achieved different results is due to the larger sampling durations and limited field trails in other work.  The sample duration limit imposed during our experiments was in line with the performance requirements of an EMV application, i.e.\ 500 milliseconds.  Additionally, transportation is one of the biggest application areas of contactless smart cards, along with banking; in this domain, the recommended duration for a transactions is far lower, in the range of 300--400 milliseconds.  Imposing a limit of 500 millisecond in our experiments, therefore, was based on the upper bound of the recommendations of two significant application areas where contactless mobile phones might be utilised. 

Proposals that suggest that a user may initiate sensor measurements in advance of the actual transaction would give us a far longer sample duration -- potentially helping in building a robust proximity detection mechanism.  We do not support this proposal at two points. Firstly, they require the user to pre-empt a transaction, which, in realistic scenarios, requires a task to be performed well in advance before they can use their mobile device.  Secondly, a user would have to start the measurement at a distance from the PT, to potentially give PI more time to measure a larger sample, which does not provide a measurement of proximity.  Furthermore, we do not subscribe to the idea that proximity detection is unnecessary because most mobile devices require a user's PIN or biometric to open a payment application.  In the relay attack variant known as a Mafia Attack, a malicious PT is deployed by an attacker to trick genuine users to tap their handset (PI) on it.  In this attack scenario, a PIN or biometric cannot protect against relay attacks.

During our experiments, we realised that sensors and their associated platforms may not have the maturity required for a wide-scale deployment as a proximity detection mechanism for NFC-enabled phones.  The variations in sensor readings across devices, how the platform's sensor architecture is affected by other applications running simultaneously, and differences in minimum sampling rates may vary across mobile device manufacturers.  We consider that mobile sensors have a considerable way to go before achieving the necessary interoperability, standardised specifications, and performance requirements to prevent relay attacks.

From the work carried out and the results presented in this paper, we can claim with a high degree of confidence that mobile sensors, at least in their current state on Google Android devices, are not suitable for use as an anti-relay mechanism.  This is especially pertinent in the case of applications with high security requirements, such as banking, transport and access-control at highly sensitive sites. It may be argued that these sensors might be suitable for low-risk application that do not have stringent transaction time limits and distance bounding assurance requirements.  However, the developer ought to consider the risks highlighted in this paper, i.e.\ EER rates at optimum thresholds.  For each sensor, we provide an EER graph that indicates the effectiveness of the respective sensor and their associated risk if it was deployed.  From the data we have analysed we can safely claim that mobile sensors, in their current state in most of the Google Android devices, may not be suitable for deployment as an anti-relay mechanism.

As part of our future research, we are currently experimenting with:

\begin{itemize}
\item Collecting and evaluation a large data set of actual relay attacks using these sensors, and investigating if and how a relay attack in the field is reflected in its sensor measurements.
\item Can we simultaneously measures multiple sensors within the transaction time duration?  If so, will it reduce the risk associated with these sensors individually?
\item Combining sensor measurements with time slices: only one sensor is measured at a time, but over the duration of the transaction multiple sensors could be used.
\item Is a location-specific deployment feasible? And will it improve the success rate significantly if such an approach is adopted?
\end{itemize}

\section{Conclusion}
%\todo[linecolor=green!90!white, backgroundcolor=purple!20!white,
 %bordercolor=red, inline]{Mainly written by Raja}
 
The aim of the paper was to evaluate and analyse a range of sensors present in modern day mobile devices, and determining which sensors, if any, would be suitable as a proximity detection mechanism in the domain of NFC mobile phone transactions.  We shortlisted 17 sensors accessible through the Google Android platform, before limiting it to those which are widely-available.  In existing literature, only five sensors have been proposed as an effective proximity detection mechanism by authors listed in Table 1. In this paper, we extend this with ten additional sensors by evaluating their suitability and effectiveness as a proximity detection mechanism on NFC-enabled mobile devices.  In total, we implemented and evaluated 15 sensors; three sensors WiFi, Bluetooth and GPS were dropped after initial tests as they were showing high sensor failure rates.  The scope of our analysis focuses on NFC-enabled mobile devices that emulate traditional smart card services, such as transportation and banking.  Any analysis or recommendation regarding these sensors is restricted to mobile contactless transactions that aim to substitute for the contactless smart card transactions in high security applications (like banking, transport and sensitive site's access control). We have neither evaluated nor claimed that similar results will be produced in other deployment scenarios of mobile sensors where distance bounding or security requirements, and transaction time limit is not as stringent as in our case (i.e traditional smart card services). 

The experimentation and analysis carried out as part of this paper showed that none of the sensors individually suitable to be deployed as an proximity detection mechanism for NFC-based mobile transactions. Finally, we will make the source code of our test-bed publicly available, along with our collected data sets, for open scrutiny and further analysis.

%\end{document}  % This is where a 'short' article might terminate
%ACKNOWLEDGMENTS are optional
%%\section{Acknowledgements}
% We appreciate the help given to us by number of people, facilitating in collecting sample measurements from students. We also thank PhD students from the Smart Card Centre and Information Security Group who gave us their valuable time during the data collection phase. Finally, we thank the Centre for Doctoral Training for funding and supporting Carlton during this work. 

%%%%% We will add the acknowledgement section back into the paper once it is published. 
{\footnotesize \bibliographystyle{acm}
\bibliography{paper}}

\begin{thebibliography}{10}

\bibitem{androidposition}
{Android API Reference Documentation: Position Sensors}.

\bibitem{AndroidAPIRef}
{Android API Reference Documentation: SensorEventListener}.

\bibitem{sensorUnits}
{Android API Reference Documentation: SensorManager}.

\bibitem{hce}
{Host-based Card Emulation}.

\bibitem{VeriFone2010}
{A Cashless Future on the Horizon}.
\newblock White paper, VeriFone, 2010.

\bibitem{transport300}
{Transit and Contactless Open Payments: An Emerging Approach for Fare
  Collection}.
\newblock White paper, Smart Card Alliance Transportation Council, November
  2011.

\bibitem{VISAMobileTicketing2013}
{The Future of Ticketing: Paying for Public Transport Journeys Using Visa Cards
  in the 21st Century}.
\newblock Whitepaper, VISA, January 2013.

\bibitem{MasterCard2014}
{MasterCard Contactless Performance Requirement}.
\newblock Online, MasterCard, March 2014.

\bibitem{Deloitte2015}
{Contactless Mobile Payments (finally) Gain Momentum}.
\newblock Online report, Deloitte, 2015.

\bibitem{EVM2015-ContactlessArchitectureReq}
Emv contactless specifications for payment systems: Book a - architecture and
  general requirements.
\newblock Specification Verstion 2.5, EMVCo, LLC, March 2015.

\bibitem{VISA-TADG}
Transactions acceptance device guide (tadg).
\newblock Specification Version 3.0, VISA, May 2015.

\bibitem{UKCardsAnnualRep2015a}
{UK Cards Annual Report 2015}.
\newblock Online report, The UK Cards Association, 2015.

\bibitem{UKCardsPayment2015a}
{UK Cards Payments 2015}.
\newblock Online report, The UK Cards Association, 2015.

\bibitem{boureanu2014towards}
{\sc Boureanu, I., Mitrokotsa, A., and Vaudenay, S.}
\newblock {Towards secure distance bounding}.
\newblock In {\em {Fast Software Encryption}\/} (2014), Springer, pp.~55--67.

\bibitem{Coskun2013}
{\sc Coskun, V., Ozdenizci, B., and Ok, K.}
\newblock {A Survey on Near Field Communication (NFC) Technology}.
\newblock {\em Wireless Personal Communications 71}, 3 (2013), 2259--2294.

\bibitem{Cremers2012}
{\sc Cremers, C., Rasmussen, K., Schmidt, B., and Capkun, S.}
\newblock {Distance Hijacking Attacks on Distance Bounding Protocols}.
\newblock In {\em {Security and Privacy (SP), 2012 IEEE Symposium on}\/} (May
  2012), pp.~113--127.

\bibitem{DrimerM07}
{\sc Drimer, S., and Murdoch, S.~J.}
\newblock {Keep Your Enemies Close: Distance Bounding Against Smartcard Relay
  Attacks.}
\newblock In {\em {USENIX Security}\/} (2007), N.~Provos, Ed., USENIX
  Association.

\bibitem{emms2014harvesting}
{\sc Emms, M., Arief, B., Freitas, L., Hannon, J., and van Moorsel, A.}
\newblock Harvesting high value foreign currency transactions from emv
  contactless credit cards without the pin.
\newblock In {\em Proceedings of the 2014 ACM SIGSAC Conference on Computer and
  Communications Security\/} (2014), ACM, pp.~716--726.

\bibitem{Ferradi2015}
{\sc Ferradi, H., Geraud, R., Naccache, D., and Tria, A.}
\newblock {When Organized Crime Applies Academic Results.}
\newblock {\em IACR Cryptology ePrint Archive\/} (2015), 20.

\bibitem{Francillon11}
{\sc Francillon, A., Danev, B., and Capkun, S.}
\newblock {Relay Attacks on Passive Keyless Entry and Start Systems in Modern
  Cars}.
\newblock In {\em {Network \& Distributed System Security}\/} (Feb. 2011),
  NDSS, The Internet Society.

\bibitem{FrancisHMM10}
{\sc Francis, L., Hancke, G.~P., Mayes, K., and Markantonakis, K.}
\newblock {Practical NFC Peer-to-Peer Relay Attack Using Mobile Phones.}
\newblock In {\em {RFIDSec}\/} (2010), S.~B.~O. Yalcin, Ed., vol.~6370 of {\em
  Lecture Notes in Computer Science}, Springer, pp.~35--49.

\bibitem{FrancisHMM11}
{\sc Francis, L., Hancke, G.~P., Mayes, K., and Markantonakis, K.}
\newblock {Practical Relay Attack on Contactless Transactions by Using NFC
  Mobile Phones.}
\newblock {\em IACR Cryptology ePrint Archive 2011\/} (2011), 618.

\bibitem{Halevi2012}
{\sc Halevi, T., Ma, D., Saxena, N., and Xiang, T.}
\newblock {Secure Proximity Detection for NFC Devices Based on Ambient Sensor
  Data}.
\newblock In {\em {Computer Security -- ESORICS 2012}}, S.~Foresti, M.~Yung,
  and F.~Martinelli, Eds., vol.~7459 of {\em {Lecture Notes in Computer
  Science}}. {Springer Berlin Heidelberg}, 2012, pp.~379--396.

\bibitem{Hancke2009615}
{\sc Hancke, G., Mayes, K., and Markantonakis, K.}
\newblock {Confidence in smart token proximity: Relay attacks revisited}.
\newblock {\em Computers \& Security 28}, 7 (2009), 615 -- 627.

\bibitem{Hancke06}
{\sc Hancke, G.~P.}
\newblock {Practical Attacks on Proximity Identification Systems (Short
  Paper).}
\newblock In {\em {IEEE Symposium on Security and Privacy}\/} (2006), IEEE
  Computer Society, pp.~328--333.

\bibitem{Hancke:2005:RDB:1128018.1128472}
{\sc Hancke, G.~P., and Kuhn, M.~G.}
\newblock {An RFID Distance Bounding Protocol}.
\newblock In {\em Proceedings of the First International Conference on Security
  and Privacy for Emerging Areas in Communications Networks\/} (Washington, DC,
  USA, 2005), SECURECOMM '05, IEEE Computer Society, pp.~67--73.

\bibitem{Hancke:2008:ATD:1352533.1352566}
{\sc Hancke, G.~P., and Kuhn, M.~G.}
\newblock {Attacks on Time-of-flight Distance Bounding Channels}.
\newblock In {\em {Proceedings of the First ACM Conference on Wireless Network
  Security}\/} (New York, NY, USA, 2008), WiSec '08, ACM, pp.~194--202.

\bibitem{scipy}
{\sc Jones, E., Oliphant, T., Peterson, P., et~al.}
\newblock {SciPy}: Open source scientific tools for {Python}, 2001--.

\bibitem{kfir2005picking}
{\sc Kfir, Z., and Wool, A.}
\newblock {Picking virtual pockets using relay attacks on contactless
  smartcard}.
\newblock In {\em {Security and Privacy for Emerging Areas in Communications
  Networks, 2005. SecureComm 2005. First International Conference on}\/}
  (2005), IEEE, pp.~47--58.

\bibitem{6378376}
{\sc Ma, D., Saxena, N., Xiang, T., and Zhu, Y.}
\newblock {Location-Aware and Safer Cards: Enhancing RFID Security and Privacy
  via Location Sensing}.
\newblock {\em Dependable and Secure Computing, IEEE Transactions on 10}, 2
  (March 2013), 57--69.

\bibitem{madlmayr2008nfc}
{\sc Madlmayr, G., Langer, J., Kantner, C., and Scharinger, J.}
\newblock {NFC devices: Security and privacy}.
\newblock In {\em {Availability, Reliability and Security, 2008. ARES 08. Third
  International Conference on}\/} (2008), IEEE, pp.~642--647.

\bibitem{markantonakis2013secure}
{\sc Markantonakis, K., and Mayes, K.}
\newblock {\em {Secure Smart Embedded Devices, Platforms and Applications}}.
\newblock Springer.

\bibitem{mehrnezhad2014tap}
{\sc Mehrnezhad, M., Hao, F., and Shahandashti, S.~F.}
\newblock {Tap-Tap and Pay (TTP): Preventing Man-In-The-Middle Attacks in NFC
  Payment Using Mobile Sensors}.
\newblock In {\em {2nd International Conference on Research in Security
  Standardisation (SSR'15)}\/} (October 2014).

\bibitem{rasmussen2010realization}
{\sc Rasmussen, K.~B., and Capkun, S.}
\newblock {Realization of RF Distance Bounding.}
\newblock In {\em {USENIX Security Symposium}\/} (2010), pp.~389--402.

\bibitem{6482441}
{\sc Roland, M., Langer, J., and Scharinger, J.}
\newblock {Applying relay attacks to Google Wallet}.
\newblock In {\em {Near Field Communication (NFC), 2013 5th International
  Workshop on}\/} (Feb 2013), pp.~1--6.

\bibitem{trujillo2010poulidor}
{\sc Trujillo-Rasua, R., Martin, B., and Avoine, G.}
\newblock {The Poulidor distance-bounding protocol}.
\newblock In {\em {Radio Frequency Identification: Security and Privacy
  Issues}}. Springer, 2010, pp.~239--257.

\bibitem{Urien201428}
{\sc Urien, P., and Piramuthu, S.}
\newblock {Elliptic curve-based RFID/NFC authentication with temperature sensor
  input for relay attacks}.
\newblock {\em Decision Support Systems 59\/} (2014), 28 -- 36.

\bibitem{Varshavsky2007}
{\sc Varshavsky, A., Scannell, A., LaMarca, A., and de~Lara, E.}
\newblock {Amigo: Proximity-Based Authentication of Mobile Devices}.
\newblock In {\em {UbiComp 2007: Ubiquitous Computing}}, J.~Krumm, G.~Abowd,
  A.~Seneviratne, and T.~Strang, Eds., vol.~4717 of {\em Lecture Notes in
  Computer Science}. Springer Berlin Heidelberg, 2007, pp.~253--270.

\end{thebibliography}
  % sigproc.bib is the name of the Bibliography in this case
% ACM needs 'a single self-contained file'!
%\balancecolumns
%\newpage
\appendix
%Appendix A
\section{Ambient Sensors}
\label{sec:AmbientSensors}
This appendix provides a short description of each sensor. %\todo[linecolor=green!80!white, backgroundcolor=yellow!40!white,bordercolor=red,inline]{Mainly written by Iakovos... Carlton to assist if required}

\subsection{Accelerometer}
The accelerometer sensor -- deployed in most modern smartphones -- measures the acceleration applied to the device on the $x$, $y$ and $z$ axes; its units are metres per second per second ($ms^{-2}$). The EER graphs based on the $MAE(R_i, M_i)$ and $corr(R_i, M_i)$ are represented in Figure~\ref{fig:EERAmbientSensor} and \ref{fig:EERAmbientSensorcorr} respectively.

\subsection{Bluetooth}
Bluetooth is a technology that facilitates wireless communication and operates in the ISM band centred at 2.4 gigahertz.  As a proximity sensor, we measure the Bluetooth devices in the vicinity (their names and MAC addresses).
%and their signal strengths.
%The EER graphs based on the $MAE(R_i, C_i)$ and $corr(R_i, C_i)$ are represented in figure {\tt{A}} and {\tt{A}} respectively.

\subsection{Geomagnetic Rotation Vector (GRV)}
The GRV sensor measures the rotation of the device using the device's magnetometer and accelerometer; it returns a vector containing the angles that the device is rotated in the $x$, $y$ and $z$ axes. The EER graphs based on the $MAE(R_i, M_i)$ and $corr(R_i, M_i)$ are represented in Figure \ref{fig:EERGRV} and \ref{fig:EERGRVcorr} respectively.

\subsection{Global Positioning System (GPS)}
The GPS sensor based a satellite-based global positioning and velocity measurement. A latitude and longitude pair is returned, representing a geographical location on Earth.
%The EER graphs based on the $MAE(R_i, C_i)$ and $corr(R_i, C_i)$ are represented in figure {\tt{A}} and {\tt{A}} respectively.

\subsection{Gravity}
The gravity sensor on mobile handsets measures the effect of Earth's gravity on the device, measured in metres per second per second ($ms^{-2}$). The EER graphs based on the $MAE(R_i, M_i)$ and $corr(R_i, M_i)$ are represented in Figure \ref{fig:EERGravity} and \ref{fig:EERGravitycoor} respectively.

\subsection{Gyroscope}
\begin{figure*}[h]
\caption{FPR and FNR Graphs}\label{fig:GRV_Results}
    \centering
\begin{xtabular*}{\textwidth}{cl}
    \begin{subfigure}[b]{0.45\textwidth}
        \includegraphics[width=\textwidth]{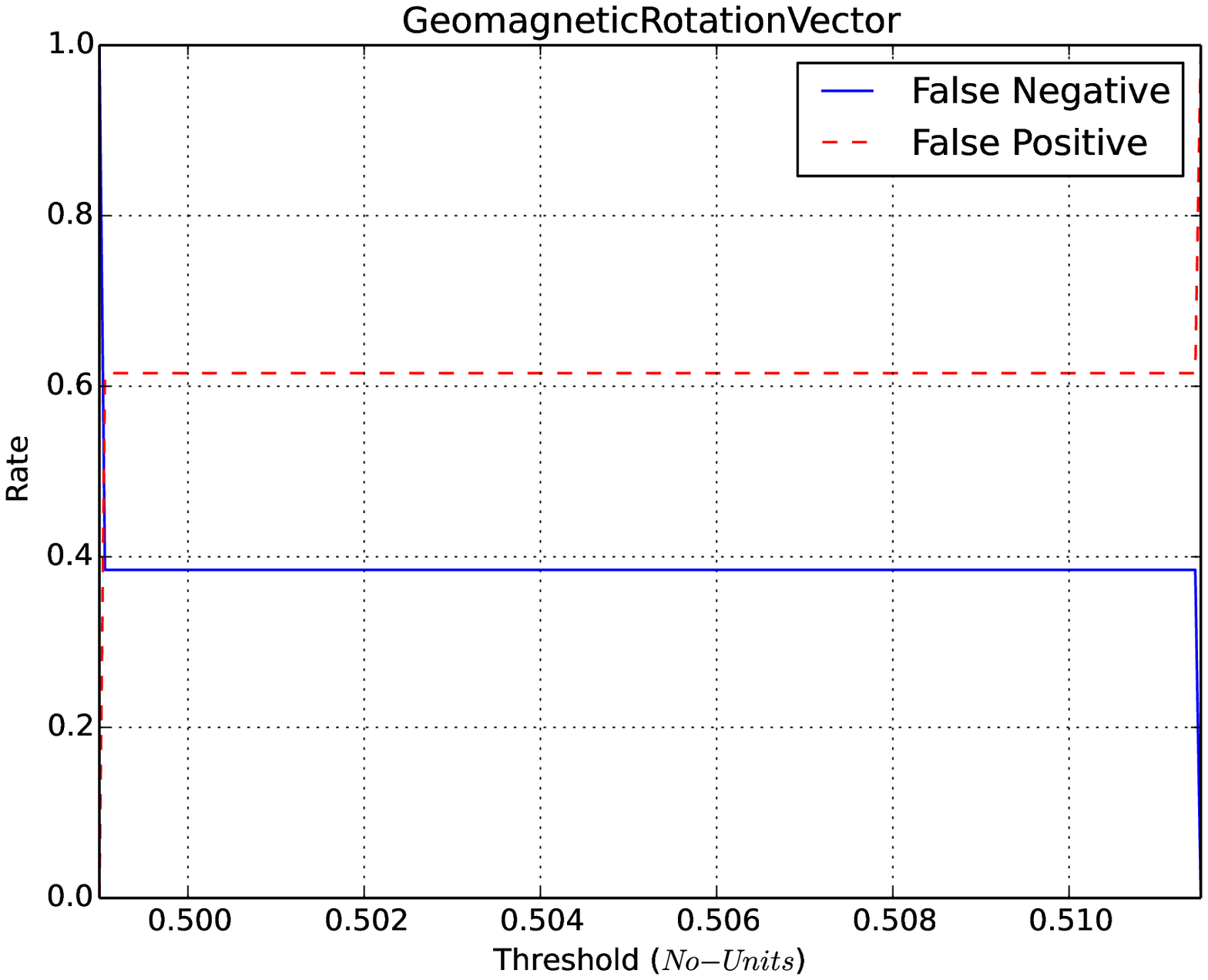}
        \caption{Based on $MAE(R_i,M_i)$}
        \label{fig:EERGRV}
    \end{subfigure}
    &
    \begin{subfigure}[b]{0.45\textwidth}
        \includegraphics[width=\textwidth]{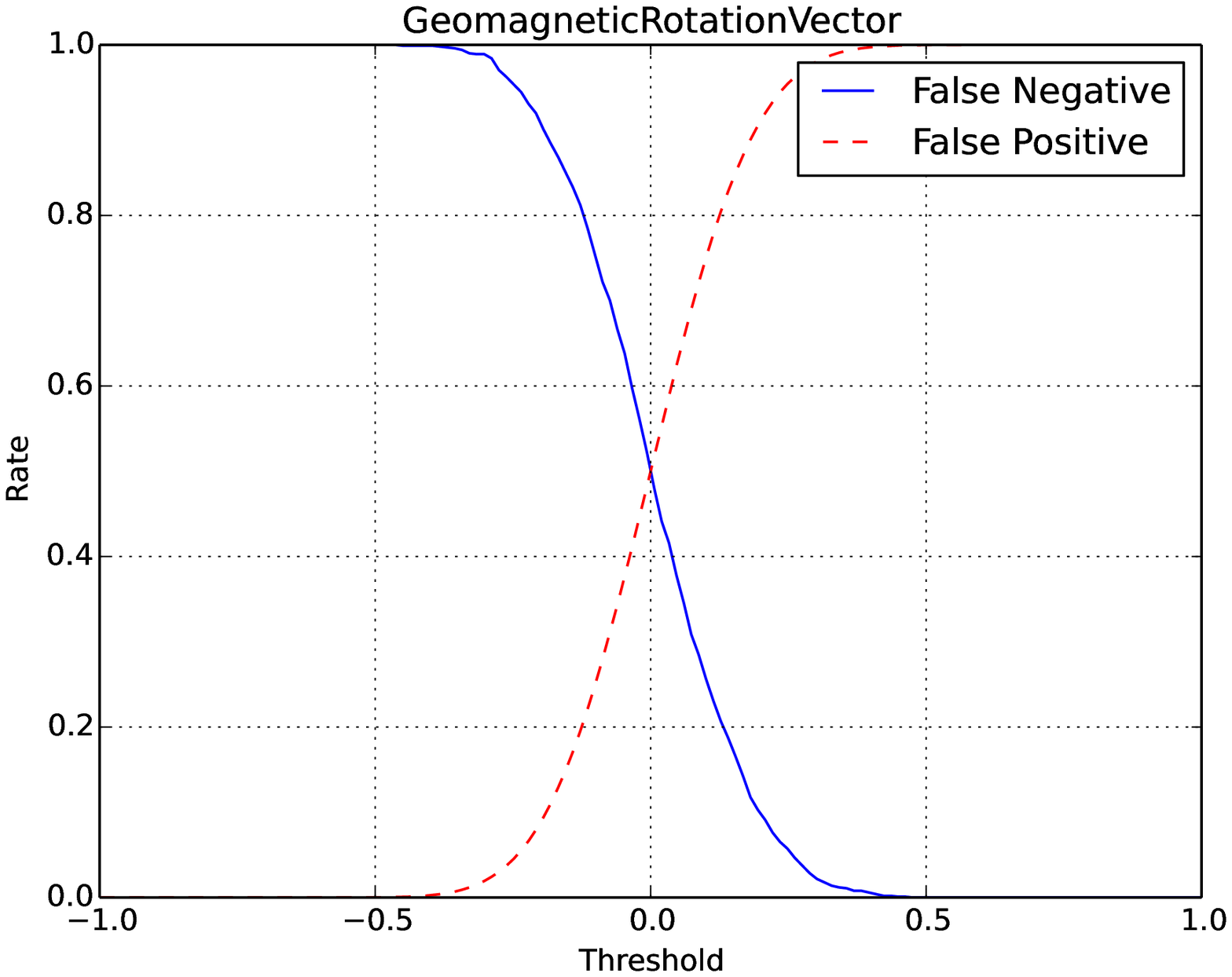}
        \caption{Based on $corr(R_i,M_i)$}
        \label{fig:EERGRVcorr}
    \end{subfigure}
 \\
    \begin{subfigure}[b]{0.45\textwidth}
        \includegraphics[width=\textwidth]{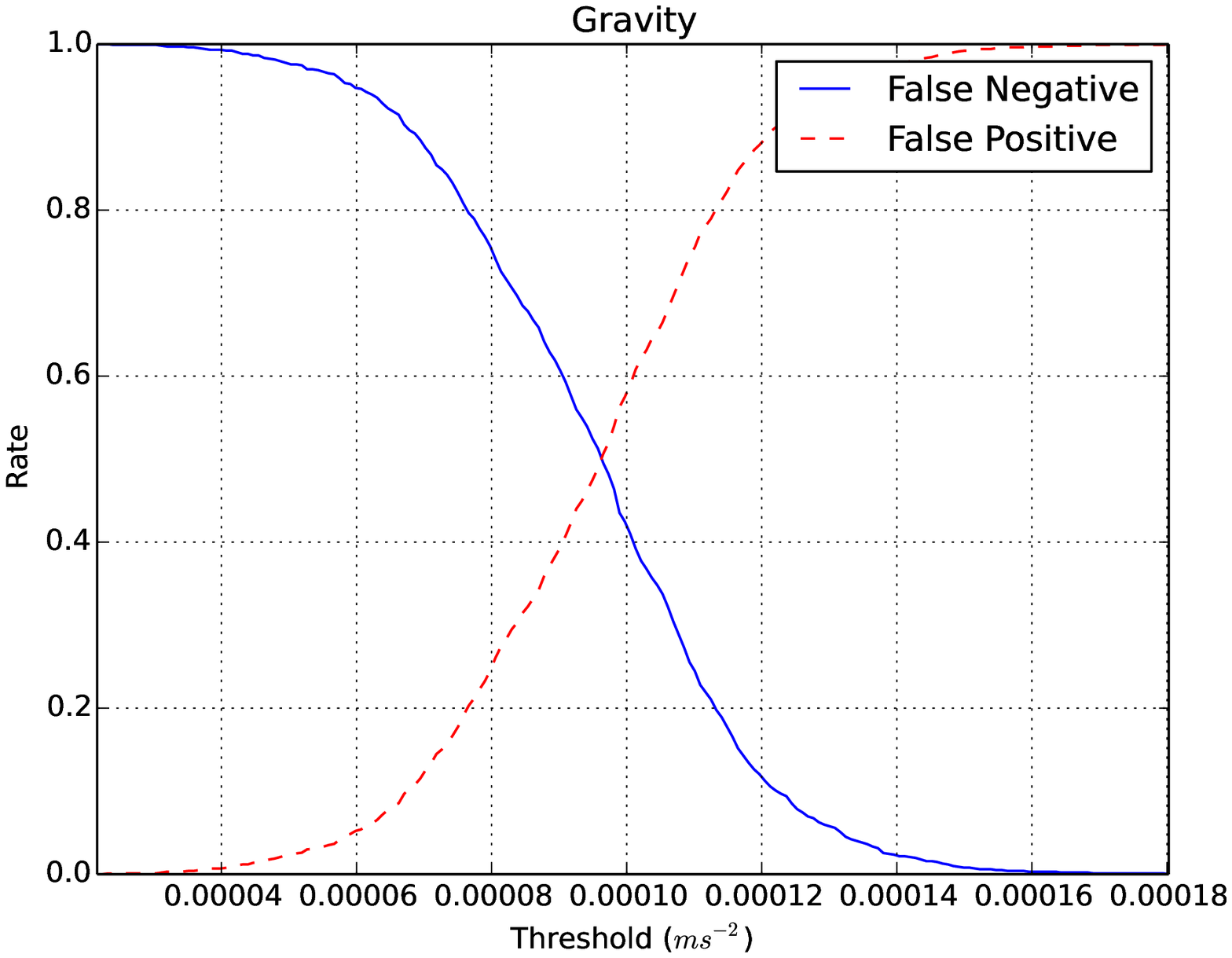}
        \caption{Based on $MAE(R_i,M_i)$}
        \label{fig:EERGravity}
    \end{subfigure}
    &
    \begin{subfigure}[b]{0.45\textwidth}
        \includegraphics[width=\textwidth]{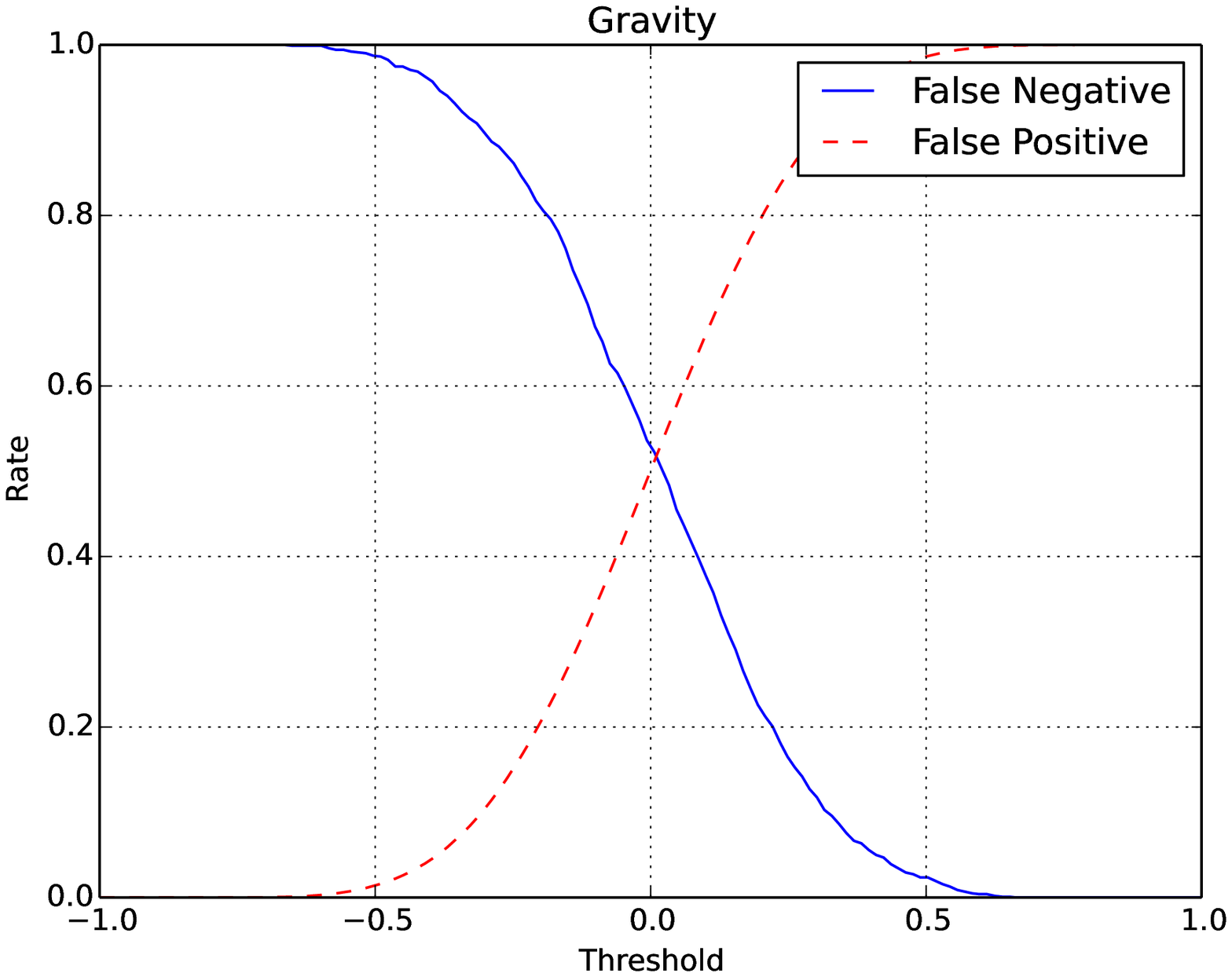}
        \caption{Based on $corr(R_i,M_i)$}
        \label{fig:EERGravitycoor}
    \end{subfigure}  
 \\
     \begin{subfigure}[b]{0.45\textwidth}
        \includegraphics[width=\textwidth]{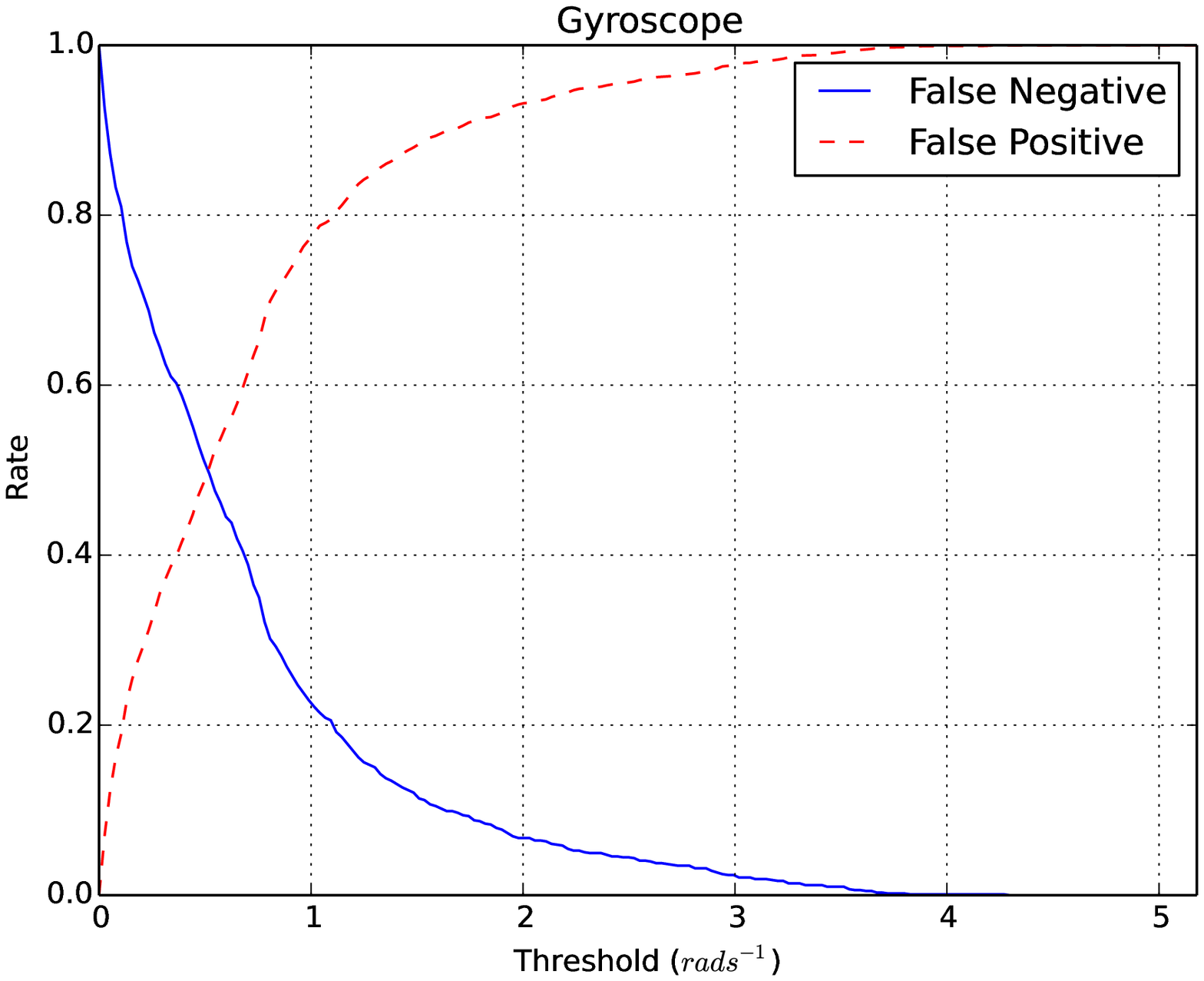}
        \caption{Based on $MAE(R_i,M_i)$}
        \label{fig:EERGyroscope}
    \end{subfigure}
    &
    \begin{subfigure}[b]{0.45\textwidth}
        \includegraphics[width=\textwidth]{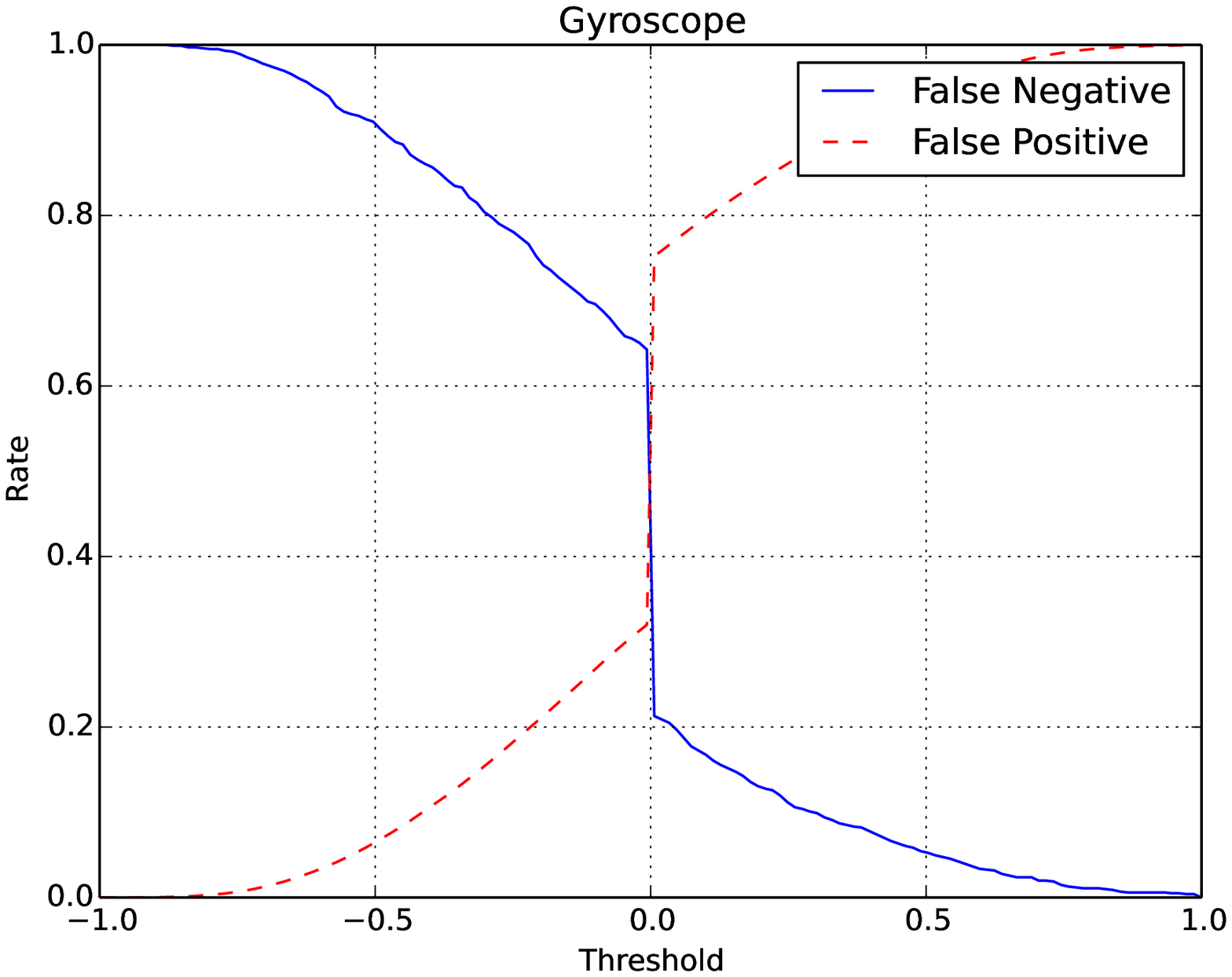}
        \caption{Based on $corr(R_i,M_i)$}
        \label{fig:EERGyroscopecoor}
    \end{subfigure}
\end{xtabular*}
\end{figure*}

The gyroscope measures the rate of rotation of the device about the $x$, $y$ and $z$ axes; its units are radians per second ($rads^{-1}$).  The EER graphs based on the $MAE(R_i, M_i)$ and $corr(R_i, M_i)$ are represented in Figure \ref{fig:EERGyroscope} and \ref{fig:EERGyroscopecoor} respectively.

\subsection{Light}
The light sensor measures the lighting conditions surrounding the mobile handset.  Android measures this quantity in $lux$.  The EER graphs based on the $MAE(R_i, M_i)$ and $corr(R_i, M_i)$ are represented in Figure \ref{fig:EER_Light} and \ref{fig:EER_Light_corr} respectively.

\subsection{Linear Acceleration}
The linear acceleration sensor measures the affect of a device's movement on itself; its units are metres per second per second ($ms^{-2}$). The EER graphs based on the $MAE(R_i, M_i)$ and $corr(R_i, M_i)$ are represented in Figure \ref{fig:EER_LinearAcceleration} and \ref{fig:EER_LinearAcceleration_corr} respectively.

\subsection{Magnetic Field}
The magnetic field sensor detects the Earth's magnetic field along three perpendicular axes $x$, $y$ and $z$.  Android measures these values in microteslas ($\mu T$).  The EER graphs based on the $MAE(R_i, M_i)$ and $corr(R_i, M_i)$ are represented in Figure \ref{fig:EER_MagneticField} and \ref{fig:EER-Corr_MagneticField_corr} respectively.  

\subsection{Network Location}
A latitude and longitude pair is returned, representing a geographical location on Earth.  The EER graphs based on the $MAE(R_i, M_i)$ is represented in Figure \ref{fig:EER_NetworkLocation}.

\subsection{Pressure}
The pressure sensor measures the atmospheric pressure surrounding the mobile handset. It is measured in hectopascals ($hPa$).
The EER graphs based on the $MAE(R_i, M_i)$ and $corr(R_i, M_i)$ are represented in Figure \ref{fig:EER_Pressure} and \ref{fig:EER-Corr_Pressure_corr} respectively.

\subsection{Proximity}
The proximity sensors detects distance, measured in centimetres.
In many devices the sensor returns only a boolean value, declaring whether something is in close proximity to the device or not.
%detects the presence of nearby objects by emitting an electromagnetic field or beam for instance infrared.
%The EER graphs based on the $MAE(R_i, C_i)$ and $corr(R_i, C_i)$ are represented in figure \ref{} and {\tt{A}} respectively.

\subsection{Rotation Vector}
Rotation vector is a software sensor, similar to the GRV, but also incorporates the gyroscope. The returned values represent the angles which the device has rotated through the $x$, $y$ and $z$ axes.
%The EER graphs based on the $MAE(R_i, C_i)$ and $corr(R_i, C_i)$ are represented in figure {\tt{A}} and {\tt{A}} respectively.

\subsection{Sound}
For the sound sensor's measurement, we use the device's microphone to record the noise in the vicinity of the mobile handsets and retrieve the maximum amplitude that was sampled, every time it becomes available by the Android operating system. The EER graphs based on the $MAE(R_i, M_i)$ and $corr(R_i, M_i)$ are represented in Figure \ref{fig:EER_Sound} and \ref{fig:EER_Sound_corr} respectively.

\subsection{WiFi}
This sensor uses traditional WiFi to detect the networks in the vicinity of the mobile device.  The MAC addresses and ESSIDs of the nearby networks are returned.
%The EER graphs based on the $MAE(R_i, C_i)$ and $corr(R_i, C_i)$ are represented in figure {\tt{A}} and {\tt{A}} respectively.

\addtocounter{figure}{-1}
\begin{figure*}[h]
    \centering
\begin{xtabular*}{\textwidth}{cl}
    \begin{subfigure}[b]{0.45\textwidth}
		\addtocounter{subfigure}{6}
        \includegraphics[width=\textwidth]{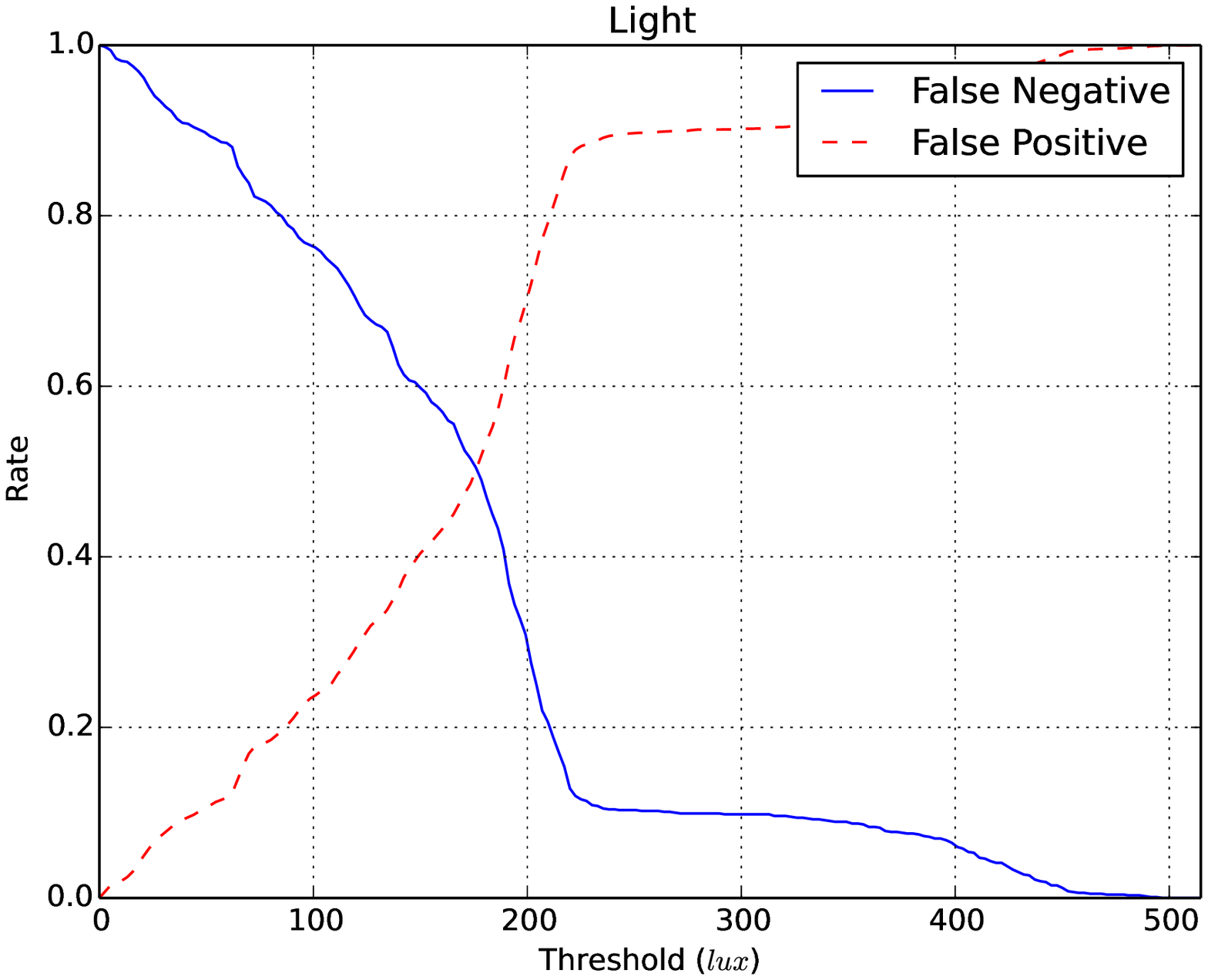}
        \caption{Based on $MAE(R_i,M_i)$}
        \label{fig:EER_Light}
    \end{subfigure}
    &
    \begin{subfigure}[b]{0.45\textwidth}
        \includegraphics[width=\textwidth]{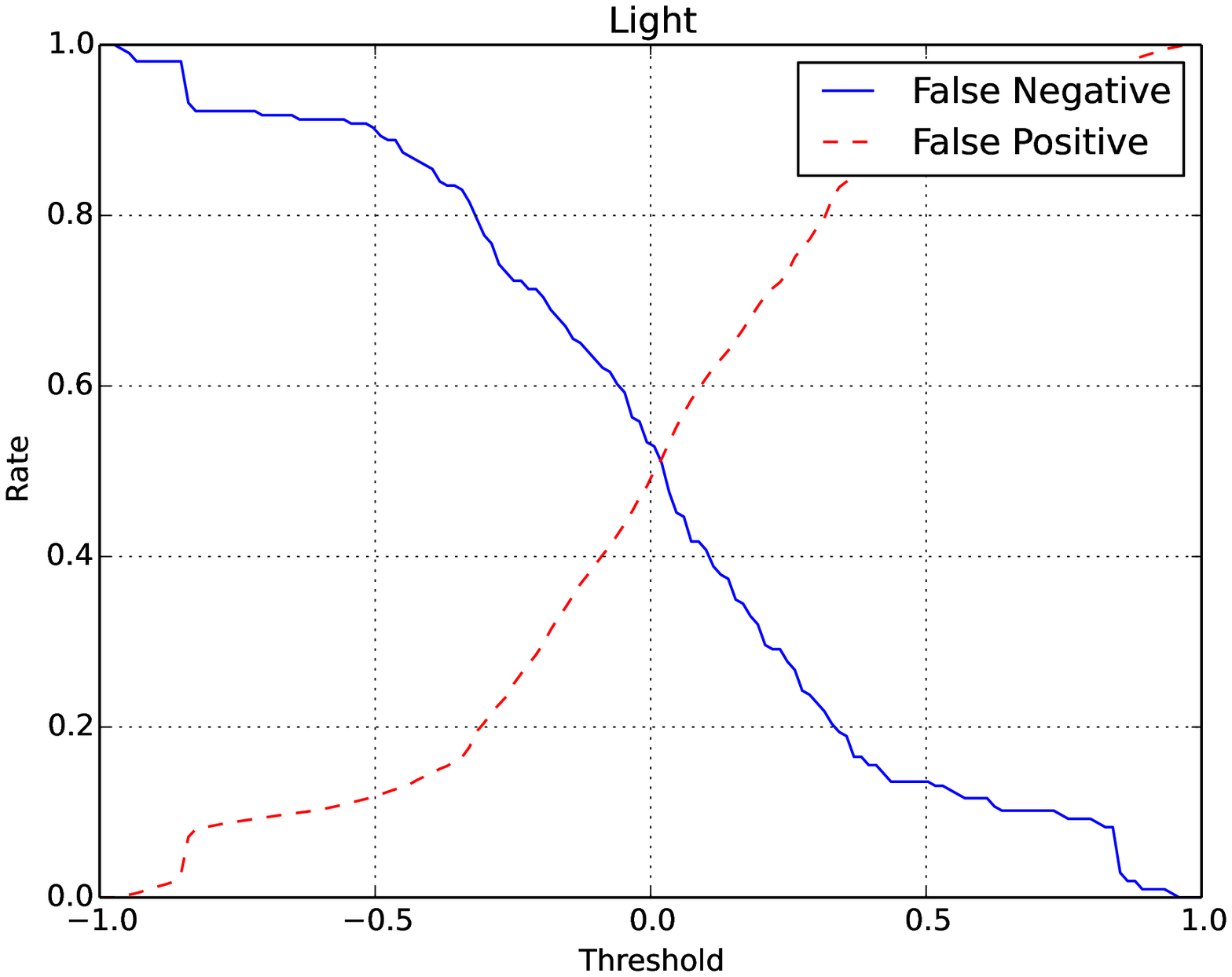}
        \caption{Based on $corr(R_i,M_i)$}
        \label{fig:EER_Light_corr}
    \end{subfigure}
\\
    \begin{subfigure}[b]{0.45\textwidth}
        \includegraphics[width=\textwidth]{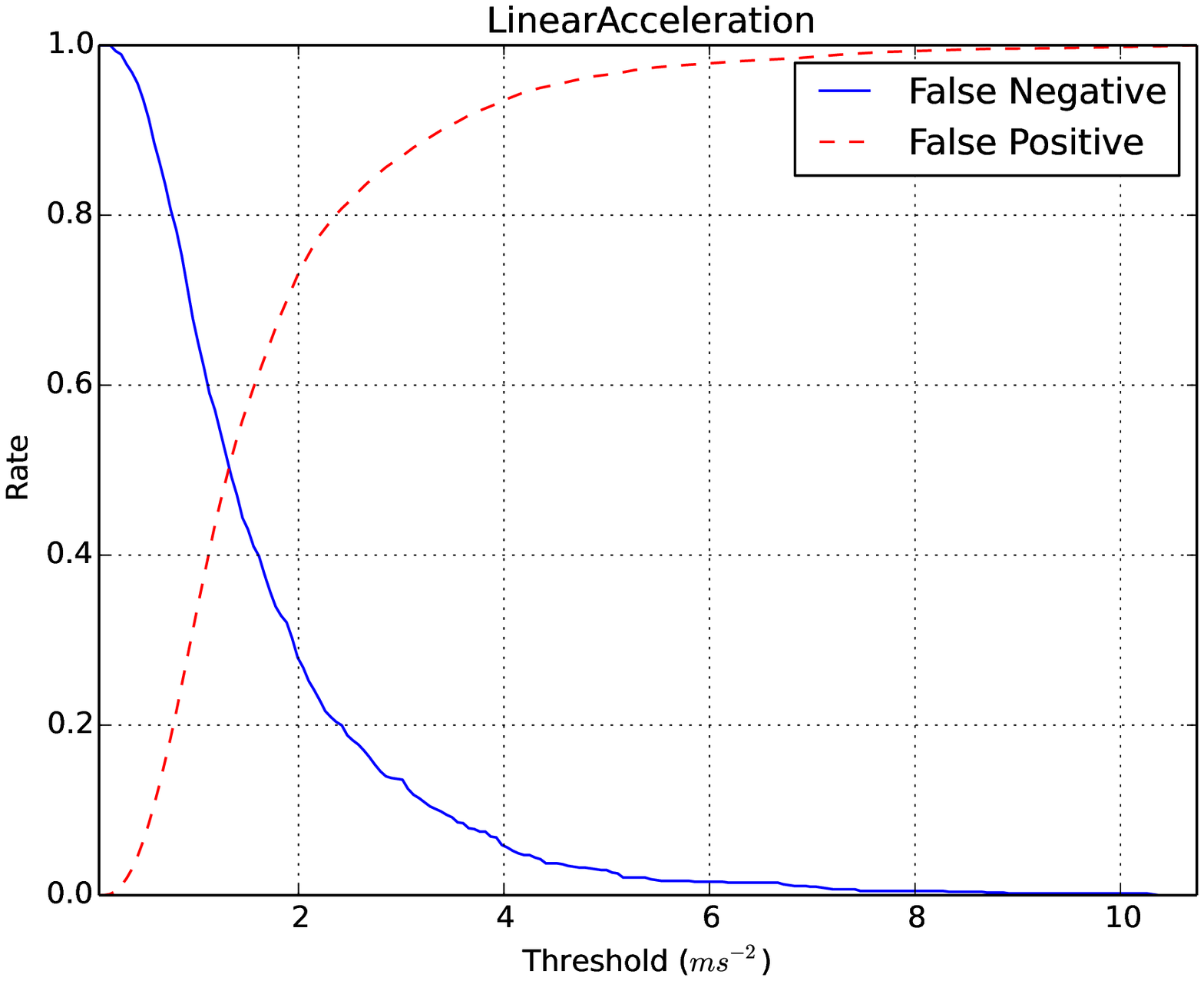}
        \caption{Based on $MAE(R_i,M_i)$}
        \label{fig:EER_LinearAcceleration}
    \end{subfigure}
    &
    \begin{subfigure}[b]{0.45\textwidth}
        \includegraphics[width=\textwidth]{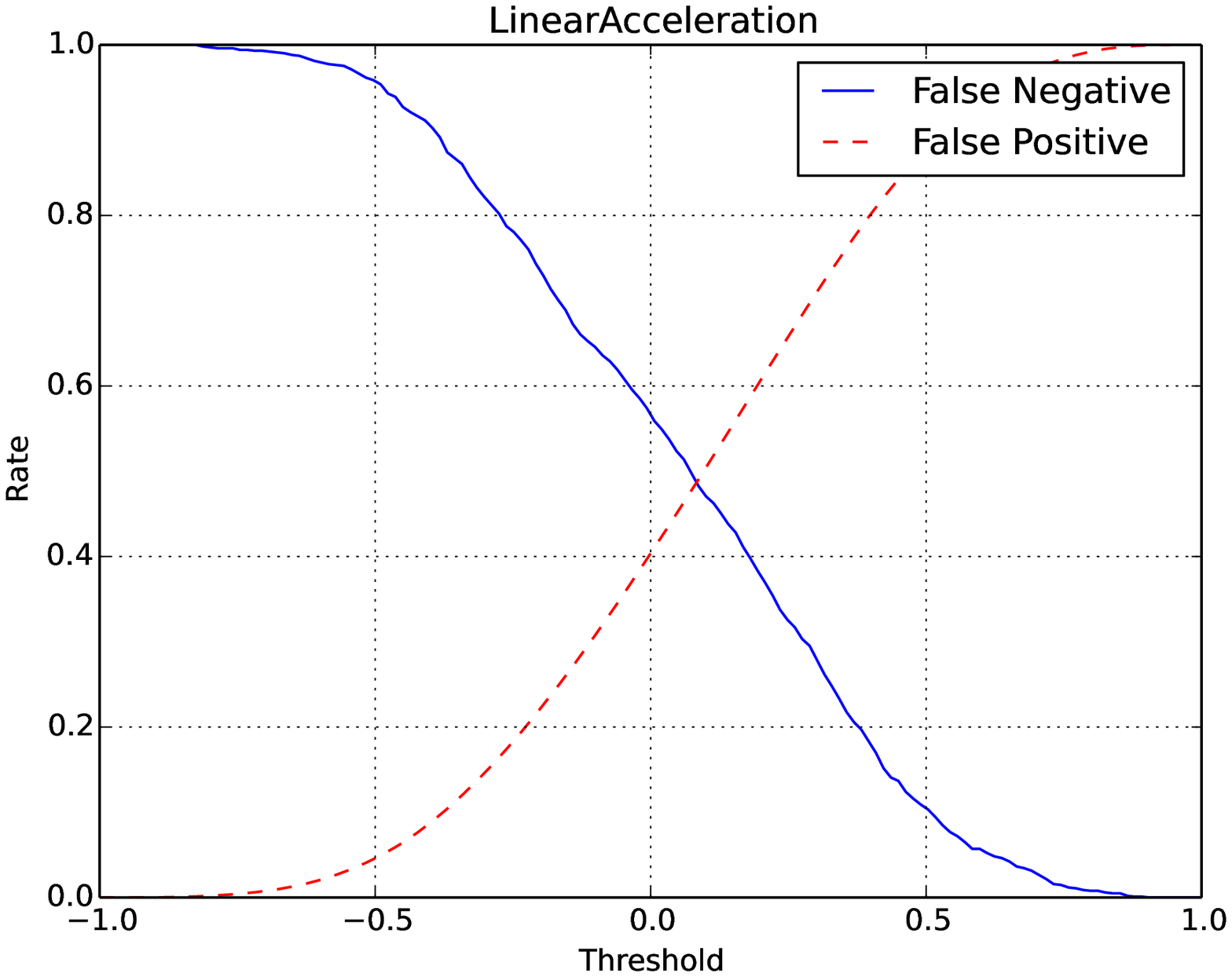}
        \caption{Based on $corr(R_i,M_i)$}
        \label{fig:EER_LinearAcceleration_corr}
    \end{subfigure}  
    \\
    \begin{subfigure}[b]{0.45\textwidth}
        \includegraphics[width=\textwidth]{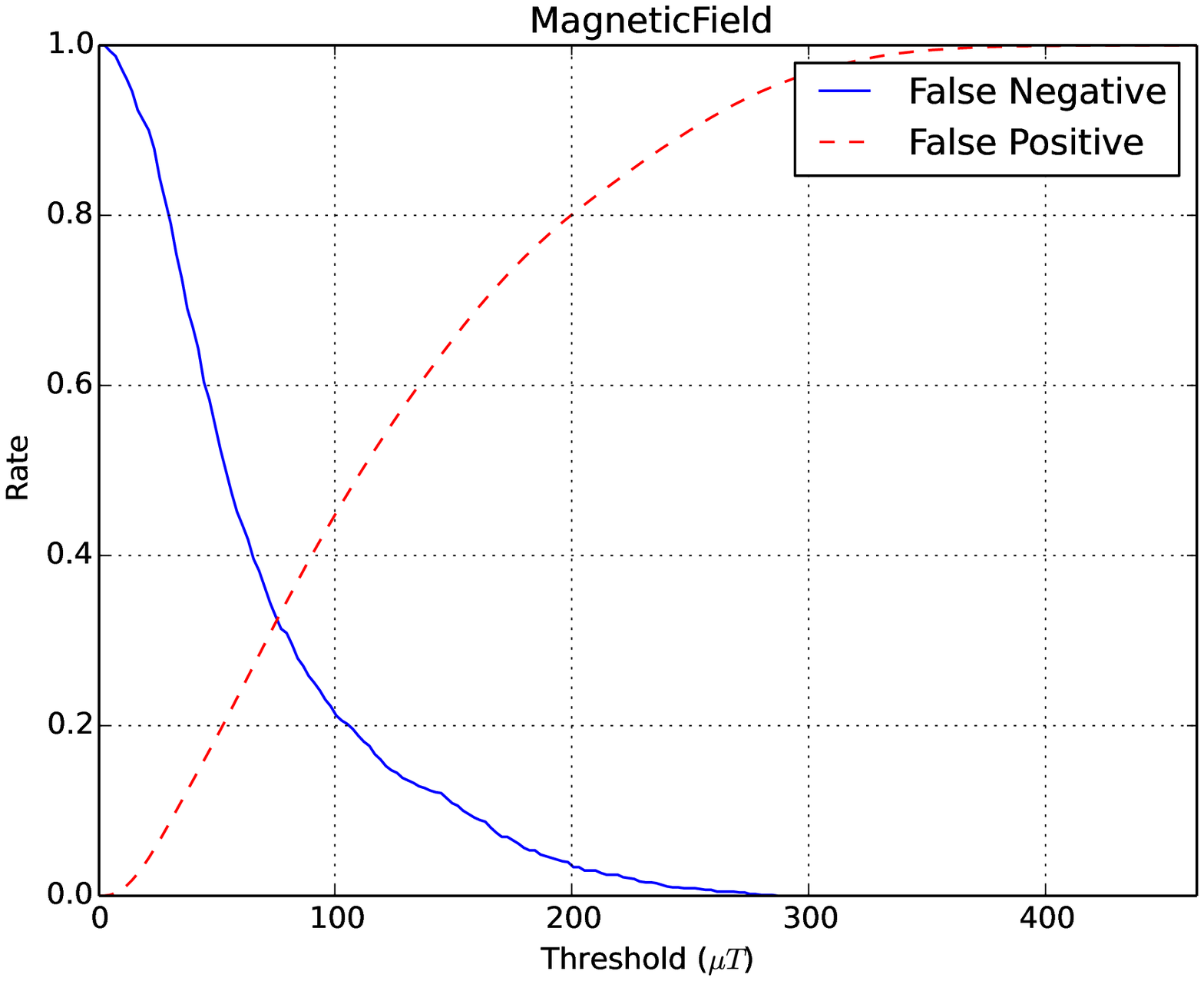}
        \caption{Based on $MAE(R_i,M_i)$}
        \label{fig:EER_MagneticField}
    \end{subfigure}
    &
    \begin{subfigure}[b]{0.45\textwidth}
        \includegraphics[width=\textwidth]{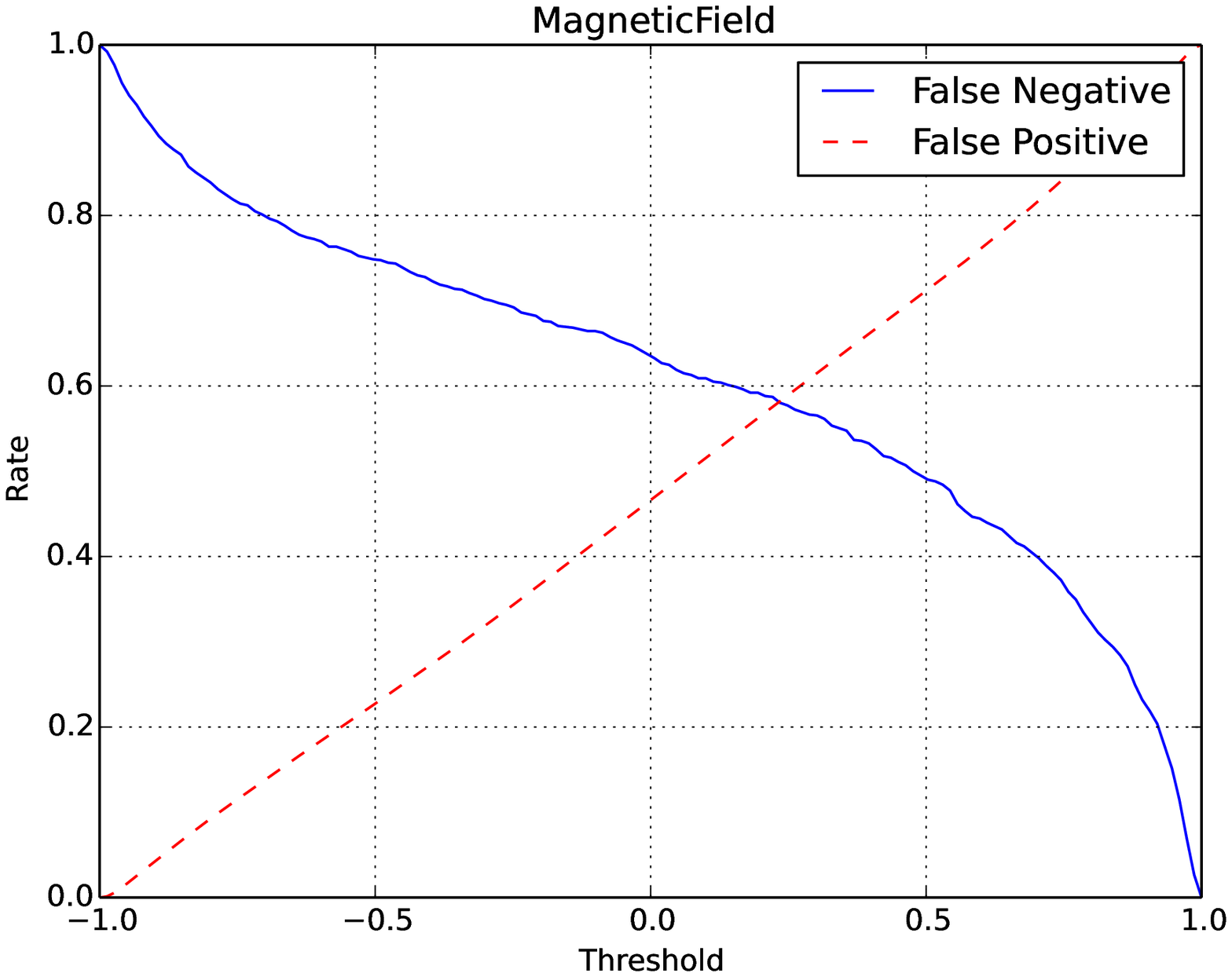}
        \caption{Based on $corr(R_i,M_i)$}
        \label{fig:EER-Corr_MagneticField_corr}
    \end{subfigure}
\end{xtabular*}
\end{figure*}

\addtocounter{figure}{-1}
\begin{figure*}[h]
    \centering
\begin{xtabular*}{\textwidth}{cl}
    \multicolumn{2}{c}{
    \begin{subfigure}[b]{0.45\textwidth}
		\addtocounter{subfigure}{12}
        \includegraphics[width=\textwidth]{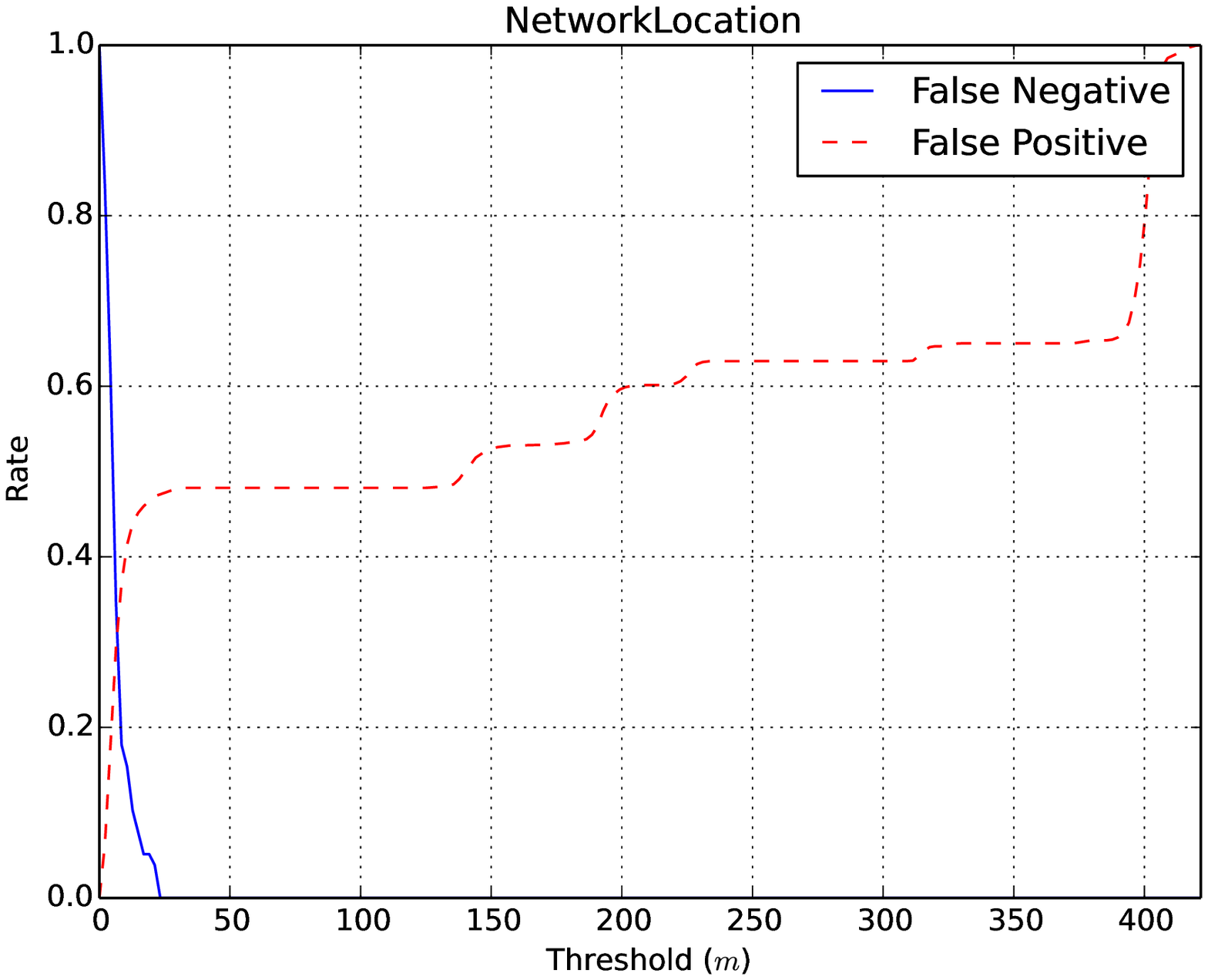}
        \caption{Based on $MAE(R_i,M_i)$}
        \label{fig:EER_NetworkLocation}
    \end{subfigure}
    }
 \\
    \begin{subfigure}[b]{0.45\textwidth}
        \includegraphics[width=\textwidth]{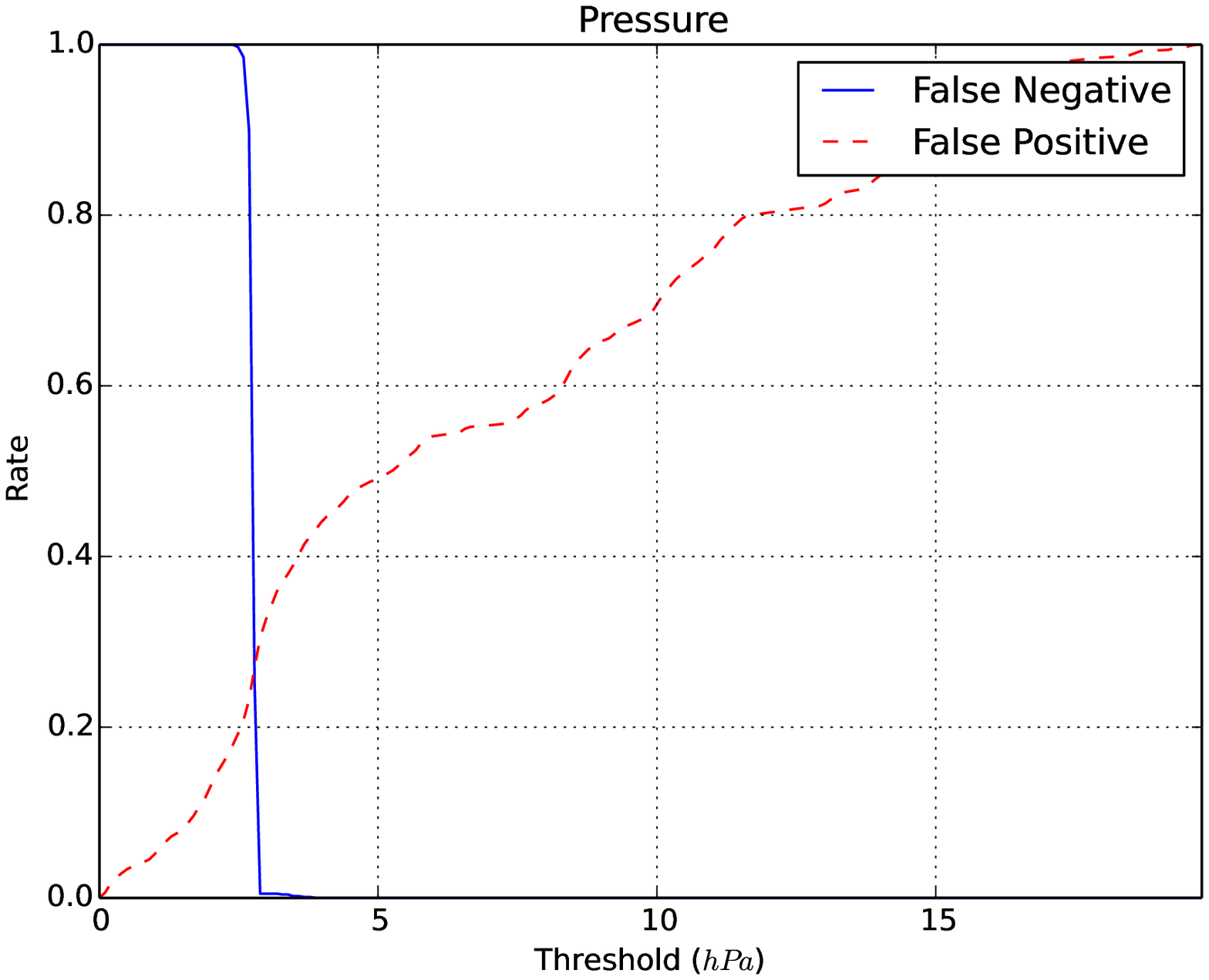}
        \caption{Based on $MAE(R_i,M_i)$}
        \label{fig:EER_Pressure}
        \end{subfigure}
 &
     \begin{subfigure}[b]{0.45\textwidth}
        \includegraphics[width=\textwidth]{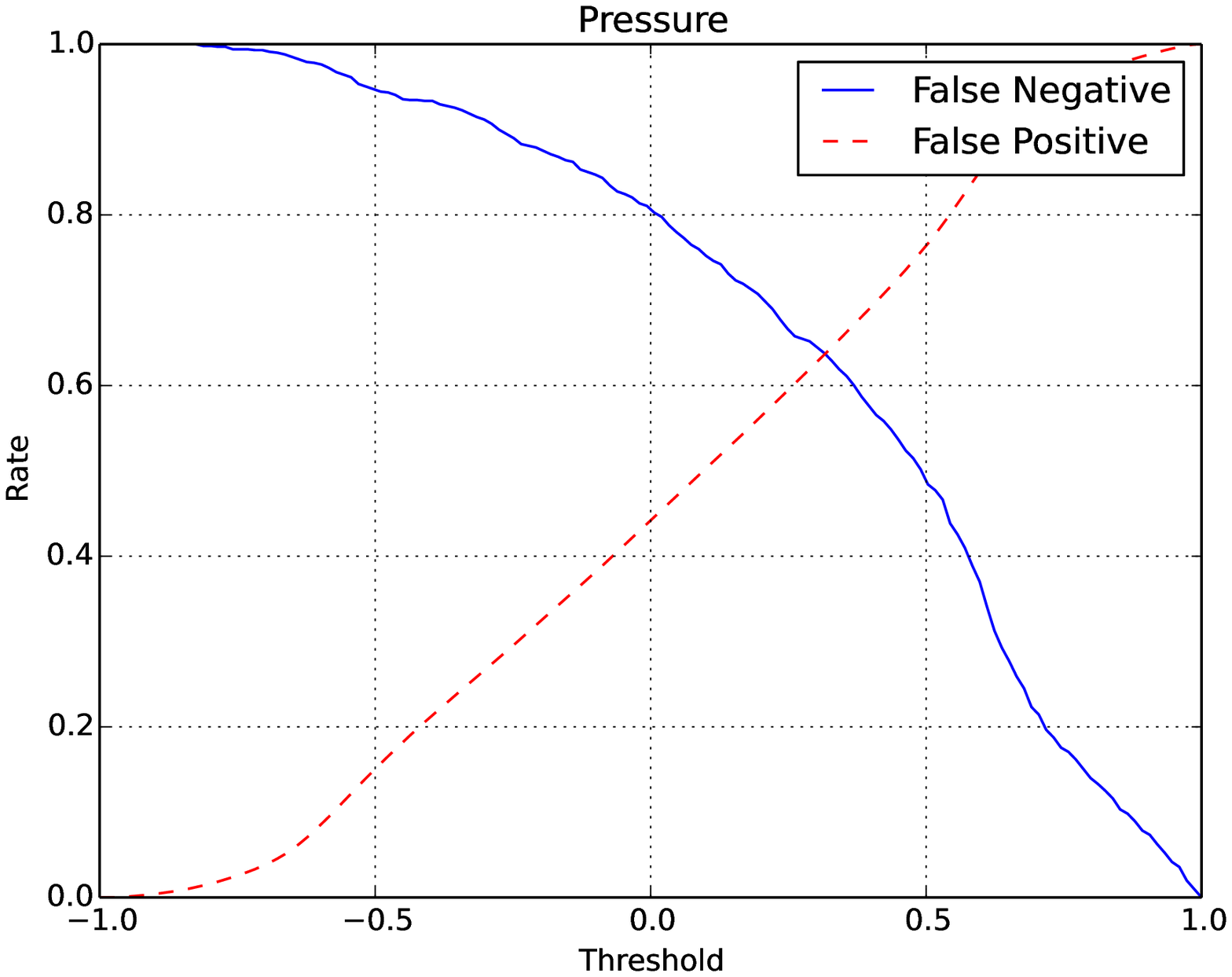}
        \caption{Based on $corr(R_i,M_i)$}
        \label{fig:EER-Corr_Pressure_corr}
    \end{subfigure}
 \\
    \begin{subfigure}[b]{0.45\textwidth}
        \includegraphics[width=\textwidth]{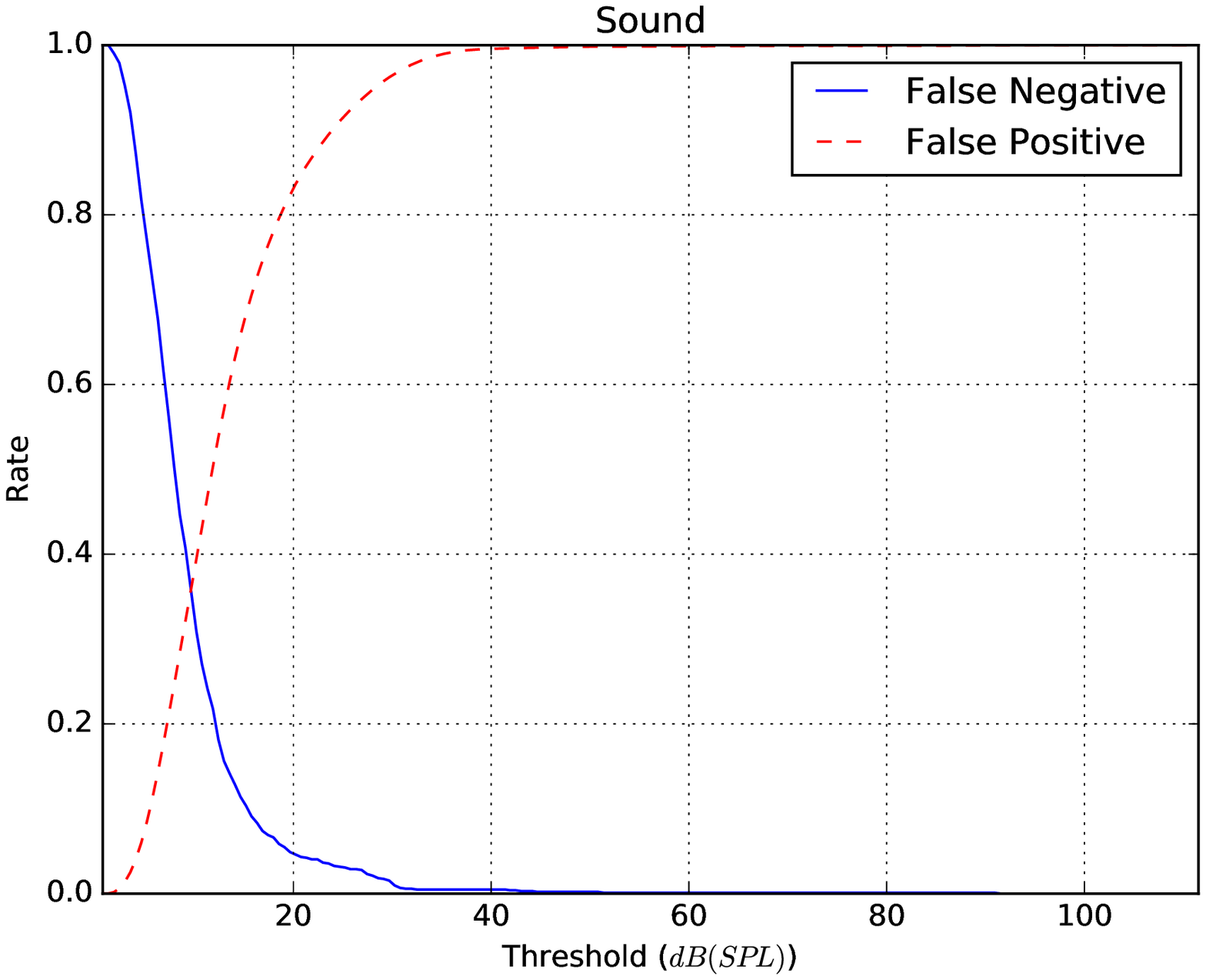}
        \caption{Based on $MAE(R_i,M_i)$}
        \label{fig:EER_Sound}
    \end{subfigure}
    &
    \begin{subfigure}[b]{0.45\textwidth}
        \includegraphics[width=\textwidth]{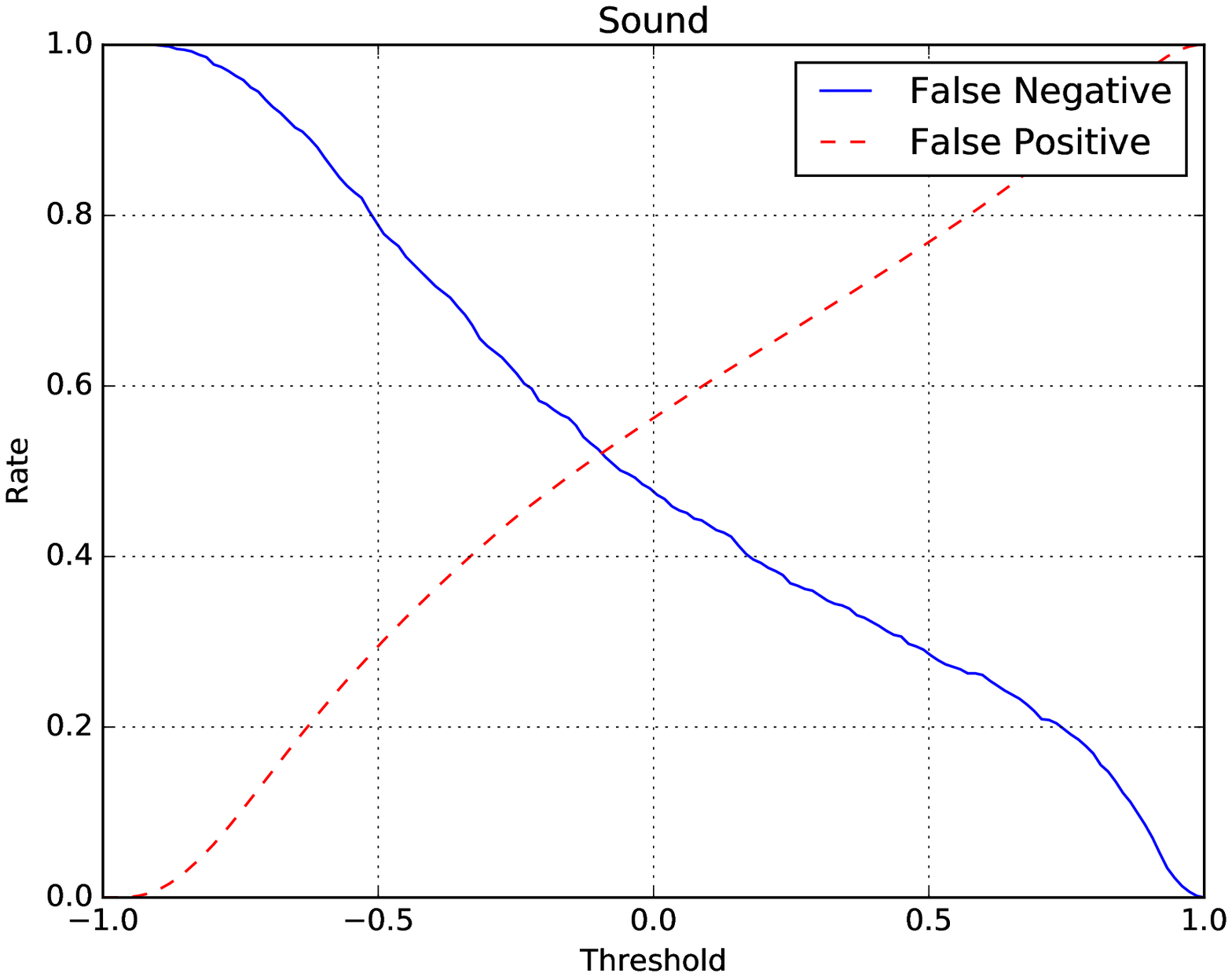}
        \caption{Based on $corr(R_i,M_i)$}
        \label{fig:EER_Sound_corr}
    \end{subfigure}
    \end{xtabular*}
  % }
    \end{figure*}

\section{Technical limitations in sensors}
\label{sec:appendixB}

\begin{itemize}
	\item \textbf{Bluetooth:} Very few times (<5\% of all measurements) data was returned in the 500ms timeframe on all devices.
    \item \textbf{Geomagnetic Rotation Vector:} Although the sensor is present on Nexus 5, no data was returned from it.
    \item \textbf{GPS:} No data is returned in the 500ms timeframe on any of the devices used in our experiments.
    \item \textbf{Light:} The Nexus 9 tablets return a limited amount of values, in large intervals.
    Although the sensor was functioning, the mobile devices were chosen for the experiments, as they produce a much wider range of values and higher accuracy.
    \item \textbf{Rotation Vector:} There were inconsistencies in the readings recorded on any of the possible pairs, therefore the sensor could not be evaluated properly.  This includes the two Nexus 9 tablets, although they were the same model and were running the same version of the operating system.  Finally, the Nexus 5 did not return any results, although the sensor is present on the device.
    \item \textbf{Sound:} No data is returned in the 500ms timeframe on the SGS5 mini.
    \item \textbf{WiFi:} No data is returned in the 500ms timeframe on any of the devices used in our experiments.
\end{itemize}

%\balancecolumns % GM June 2007
% That's all folks!
\end{document}